\documentclass[11pt]{article}
\pdfoutput=1

\usepackage[usenames]{xcolor}
\usepackage[english]{babel}
\usepackage[small]{caption}
\usepackage{verbatim}
\usepackage{multirow}
\usepackage{relsize}
\usepackage{amsmath,amssymb,amsbsy,amstext,amsfonts}
\usepackage{epsfig,multicol,bbm,simplewick}
\usepackage{mathtools}
\usepackage{enumitem}
\usepackage{csquotes}
\usepackage{tensor}
\usepackage{fix-cm}
\usepackage{booktabs}
\usepackage{graphicx}
\usepackage{bm}
\usepackage{subfigure}
\usepackage{float}
\usepackage[explicit]{titlesec}
\usepackage{enumitem}
\usepackage{setspace}
\usepackage{blindtext}
\usepackage{textpos}
\usepackage{etoolbox}
\usepackage[linktocpage=true]{hyperref}
\usepackage{anyfontsize}
\usepackage{listings,xfp}
\usepackage[dotinlabels]{titletoc}
\usepackage[normalem]{ulem}
\usepackage{esvect}
\DeclareMathAlphabet\mathbfcal{OMS}{cmsy}{b}{n}
\usepackage{tikz}
\usepackage{tikz-feynman}
\usetikzlibrary{patterns.meta}

\usepackage[maxbibnames=99,style=numeric-comp,sorting=none,doi=false,isbn=false]{biblatex}
\AtEveryBibitem{\clearfield{pages}} \DeclareFieldFormat[article,misc,book]{title}{\mkbibbold{#1}}
\renewbibmacro{in:}{}
\definecolor{mcoolred}{rgb}{1, 0.2, 0.34}
\hypersetup{colorlinks=true,linkcolor=blue,citecolor=blue,urlcolor=blue}
\usepackage[a4paper, total={6.75in, 9.75in}]{geometry}
\newcommand\numberthis{\addtocounter{equation}{1}\tag{\theequation}}
\renewcommand{\thesection}{\arabic{section}}

\dottedcontents{section}[1.2em]{}{1.2em}{6pt}

\titleformat{\section}[block]
  {\titlerule\addvspace{4pt}\normalfont\fontsize{14}{16}\bfseries}
  {\thesection.\enspace}{0pt}{\large #1}[\vspace{2pt}\titlerule]

\titleformat{\subsection}[block]
  {\addvspace{4pt}\normalfont\fontsize{14}{16}\bfseries}
  {$\blacksquare$\enspace\thesubsection.\enspace}{0pt}{\large #1}[\vspace{2pt}\titlerule]

\makeatletter
\def\@xfootnote[#1]{%
  \protected@xdef\@thefnmark{#1}%
  \@footnotemark\@footnotetext}
\makeatother

\begin{document}

\begin{titlepage}

\setcounter{page}{1} \baselineskip=15.5pt \thispagestyle{empty}

\hfill DESY-24-149

\begin{center}
{\fontsize{16}{30}\selectfont \bf 
Gravitational waves in ultra-slow-roll
\vspace{0.1in}\\
and their anisotropy at two loops}
\end{center}

\begin{NoHyper}
\begin{center}
{\fontsize{14}{30}\selectfont
Juan \'Alvarez Ruiz$^{1}$\footnote[]{juan.alvarez.ruiz@rai.usc.es}
and 
Juli\'an Rey$^{2}$\footnote[]{julian.rey@desy.de}}
\end{center}
\end{NoHyper}

\begin{center}
\textsl{
$^{1}$Universidade de Santiago de Compostela, E-15782 Galicia-Spain
\\
$^{2}$Deutsches Elektronen-Synchrotron DESY, Notkestr. 85, 22607 Hamburg, Germany
}

\end{center}

\vspace{0.6cm}
\hrule
\vspace{0.4cm}
\noindent {\large \textbf{Abstract}} \\[0.3cm]
\begin{onehalfspacing}
\noindent We compute the non-Gaussian corrections to the energy density and anisotropies of gravitational waves induced during the radiation era after an ultra-slow-roll phase of inflation by using a diagrammatic approach, and present the corresponding Feynman rules. Our two-loop calculation includes both the intrinsic non-Gaussianity of the inflaton perturbation $\delta\phi$ and the non-Gaussianity arising from the nonlinear relation between the latter and the curvature perturbation $\mathcal{R}$, which we find to be subdominant with respect to the former. We apply our formalism to an analytical model in which the ultra-slow-roll phase is followed by a constant-roll stage with a nonvanishing second slow-roll parameter $\eta$, and address the renormalization of the one-loop scalar power spectrum in this scenario.
\par
\end{onehalfspacing}
\vspace{0.4cm}
\hrule

\vspace{0.1in}
\tableofcontents
\vspace{0.2in}
\titlerule

\end{titlepage}

\phantomsection
\section*{Introduction}
\addcontentsline{toc}{section}{Introduction}

Despite having been studied for about four decades at the time of writing \cite{PhysRevD.23.347}, inflation remains a theoretically rich playground for quantum fields in curved spacetimes. Significant efforts have been devoted in recent years to the study of the theory beyond the slow-roll paradigm, in regimes in which much of the standard lore may break down (see e.g.\,\cite{Celoria:2021vjw}). In particular, an ultra-slow-roll (USR) phase in which the field slows down and the scalar power spectrum is enhanced by several orders of magnitude leads to large non-Gaussianities that can significantly alter the probability distribution function of the comoving curvature perturbation $\mathcal{R}$, in stark contrast to the Gaussianity inherent to the slow-roll scenario \cite{Figueroa:2020jkf,Taoso:2021uvl,Ferrante:2022mui,Ballesteros:2024pwn,Caravano:2024moy,Ballesteros:2024pbe}. These large scalar fluctuations also act as a source for tensor modes at second order in perturbations \cite{Tomita:1967wkp,Matarrese:1992rp,Matarrese:1993zf,Carbone:2004iv,Ananda:2006af,Baumann:2007zm}, generating a peak in the gravitational wave spectrum which is potentially within reach of future space-based interferometers \cite{Bartolo:2018evs,Bartolo:2018rku,LISACosmologyWorkingGroup:2023njw}. The goal of this work is to determine how the aforementioned non-Gaussianity of the scalar modes affects this gravitational wave signal.\footnote{See \cite{Balaji:2022dbi} for work in a similar direction.} In this work, we focus only on gravitational waves induced during the radiation era, although this process can also take place during inflation, see e.g.\,\cite{Fumagalli:2021mpc,Firouzjahi:2023btw,Ballesteros:2024qqx}.

Schematically, at second order in perturbations the amplitude of the transverse-traceless tensor modes is $h\sim \mathcal{R}^2$. The energy density of gravitational waves can be obtained by computing the power spectrum $\langle h^2\rangle\sim \langle \mathcal{R}^4\rangle$. If the fluctuations obey Gaussian statistics, this four-point function reduces to the product of two scalar power spectra, $\langle \mathcal{R}^4\rangle\sim \langle\mathcal{R}^2\rangle^2$, but in the presence of non-Gaussianity more terms are possible. A popular ansatz consists of assuming these non-Gaussianities are of the local form, with $\mathcal{R}=\mathcal{R}_{\rm G}+f_{\rm NL}\mathcal{R}_{\rm G}^2$ for some Gaussian variable $\mathcal{R}_{\rm G}$, which would lead, for instance, to additional terms of the form $\langle \mathcal{R}^4\rangle\sim \langle\mathcal{R}_{\rm G}^2\rangle^2+f_{\rm NL}^2\langle\mathcal{R}_{\rm G}^2\rangle^3+\cdots$. The effect of such terms on the gravitational wave spectrum has been extensively studied \cite{Garcia-Bellido:2017aan,Cai:2018dig,Unal:2018yaa,Bartolo:2019zvb,Atal:2021jyo,Adshead:2021hnm,Ragavendra:2021qdu,Garcia-Saenz:2022tzu,Abe:2022xur,Li:2023qua,Li:2023xtl,Papanikolaou:2024kjb,Perna:2024ehx}. Non-Gaussianities generated during an ultra-slow-roll phase are only partially of the local form, however \cite{Taoso:2021uvl}. To see this, let us note that the calculation of inflationary correlators is most easily performed in the so-called $\delta\phi$ gauge (where $\delta\phi$ denotes the inflaton perturbation), in which the spatial part of the metric takes a particularly simple form. In this gauge, one can show that the only relevant terms during an ultra-slow-roll phase are the self-interactions of the inflaton field arising from the expansion of the potential, $\partial_\phi^n V\delta\phi^n$ \cite{Ballesteros:2024zdp}. The effect of these interactions on the correlators $\langle\delta\phi^n\rangle$ must be calculated perturbatively using the in-in formalism. Since the variable of interest is not $\delta\phi$ but $\mathcal{R}$, a gauge transformation of the form $\mathcal{R}\sim\delta\phi + f_{\rm NL}\delta\phi^2+\cdots$ must be performed. If the self-interactions of the field are negligible, then $\delta\phi$ is a Gaussian variable and this relation leads to non-Gaussianities of the local form, but, as we will show, this is not necessarily the case.

Diagrammatically, the spectrum of scalar-induced gravitational waves is represented by a one-loop graph at leading order (since the tensor modes are free at the linear level, in the absence of anisotropic stress), and the effects of non-Gaussianities correspond to graphs of higher order in loops. An issue that arises immediately when one attempts to compute the gravitational wave spectrum at two loops is that of renormalizing the scalar power spectrum. The calculation of loop corrections to the scalar two-point function in the presence of a USR phase has recently been the subject of intense debate, see e.g.\,\cite{Kristiano:2022maq,Riotto:2023hoz,Firouzjahi:2023aum,Firouzjahi:2023ahg,Franciolini:2023agm,Fumagalli:2023hpa,Maity:2023qzw,Ballesteros:2024zdp,Inomata:2024lud,Fumagalli:2024jzz} and the references therein. It has been shown that, depending on the duration of the ultra-slow-roll phase and how smoothly the field transitions in and out of it, the loop corrections to the scalar power spectrum around the peak can be significant, and perturbation theory effectively breaks in a certain region of parameter space \cite{Franciolini:2023agm,Ballesteros:2024zdp}.\footnote{One point of contention in the debate is whether the presence of the USR phase can influence what happens on CMB scales. Throughout this work we are interested only in accurately predicting the gravitational wave spectrum around the peak, so this point of the discussion is immaterial to us. Nonetheless, let us point out that we largely follow the treatment of \cite{Ballesteros:2024zdp}, in which loop corrections are shown to have the same scale-dependence as the tree-level spectrum on CMB scales, so that both contributions are indistinguishable from each other and the large-scale spectrum is therefore unaffected by the presence of the peak.} The interactions that control the size of these loop corrections are, in fact, the very same that influence the size of the non-Gaussianities, so the two questions are closely tied together. In practical terms, several of the diagrams that arise when calculating the tensor power spectrum at two loops involve replacing one of the scalar propagators by its loop-corrected version. Naturally, this means that perturbation theory can break for the tensor two-point function in the same way as it does for its scalar counterpart, as we will show.

Determining the origin of a stochastic gravitational wave background is a difficult task, since many models can produce spectra with similarly peaked shapes \cite{Caprini:2018mtu}. Thus, in discriminating between these models it becomes necessary to explore other observables. A compelling possibility is to study the anisotropies in this background. We expect any stochastic gravitational wave background to be anisotropic to some extent, as it happens for the cosmic microwave background. Cosmological sources typically produce much smaller anisotropies than astrophysical ones,\footnote{See e.g.\,\cite{Bodas:2022urf} for an estimate of the anisotropies in the context of cosmological phase transitions, and \cite{NANOGrav:2023tcn} for an estimate of the anistotropies produced by a background of supermassive black hole binaries.} and inflation is no exception. Since the generation of scalar-induced gravitational waves is a local process, widely separated patches in the sky (which correspond to low multipoles and are therefore easier to detect \cite{Bartolo:2019zvb}) are essentially uncorrelated if the fluctuations are Gaussian. This conclusion changes in the presence of non-Gaussianity, however \cite{Bartolo:2019oiq,Bartolo:2019zvb,Bartolo:2019yeu,Dimastrogiovanni:2022afr}. If, for instance, we have non-Gaussianity of the local form $\mathcal{R}\sim f_{\rm NL}\delta\phi^2$, then in Fourier space the relation turns into a convolution and the possibility that two short-wavelength modes conspire to create a long-wavelength one opens up, effectively correlating distant patches. Moreover, since non-inflationary cosmological sources typically produce stochastic gravitational wave backgrounds that are generically Gaussian in character \cite{Bodas:2022urf}, these anisotropies could help distinguish scalar-induced gravitational waves from other types of signals.

The paper is structured as follows. In Section \ref{sec:action} we determine the interactions relevant for the calculation of the two-loop gravitational wave spectrum and address the issue of computing the one-loop corrections to the scalar power spectrum, effectively generalizing the results of \cite{Ballesteros:2024zdp} to the case in which the ultra-slow-roll phase is followed by a constant-roll stage with non-vanishing second slow-roll parameter. In Section \ref{sec:waves}, we calculate the gravitational wave energy density at two loops using a diagrammatic approach, extending the formalism of \cite{Li:2023xtl,Bartolo:2019zvb,Bartolo:2019yeu} by including the intrinsic non-Gaussianity of the inflaton field calculated using the in-in formalism, as opposed to a local ansatz. In Section \ref{sec:anis} we review the line-of-sight formalism for the computation of the anisotropies and establish the role of the non-Gaussianities in the calculation of the latter. Finally, in Section \ref{sec:numerical} we study the structure of the divergences in the diagrams and provide our numerical results. We find that, depending on the duration and smoothness of the transitions in and out of USR, perturbation theory can be violated in the gravitational wave sector as is the case with the scalar modes \cite{Ballesteros:2024zdp}. Moreover, the dominant contribution in the corrections to both the gravitational wave spectrum and the anisotropies arises from the intrinsic non-Gaussianity of the field, and therefore cannot be captured by a local ansatz for the non-Gaussianity.

\section{Action for the fluctuations}
\label{sec:action}

In this Section we provide the action for the inflationary fluctuations in the presence of a USR phase. We work in the flat gauge, where the relevant interactions for the inflaton field fluctuations are more easily obtained, and perform a gauge transformation to obtain the curvature perturbation. We then determine the diagrams contributing to the power spectrum at one loop, which are also necessary for the evaluation of the gravitational wave correlators.

\subsection{Relevant interactions}

Our setup consists of an inflaton field $\Phi$ minimally coupled to gravity
\begin{equation}
S
=
\int d^4x\sqrt{-g}
\bigg[
\frac{1}{2}R-\frac{1}{2}g^{\mu\nu}\partial_\mu\partial_\nu\Phi-V(\Phi)
\bigg].
\end{equation}
We work in natural units and set $M_p=1$. We perturb the metric and the field around time-dependent background values and expand the action in powers of the fluctuations order by order. Following \cite{Maldacena:2002vr}, the calculation is most easily performed using the ADM formalism, where the metric is parameterized as
\begin{equation}
ds^2
=
-N^2dt^2+h_{ij}(dx^i+N^idt)(dx^j+N^jdt).
\end{equation}
Here, the lapse $N$ and shift $N^i$ act as Lagrange multipliers whose equations of motion must be solved and plugged back into the action.

We work in the flat (or $\delta\phi$) gauge, defined by
\begin{equation}
\Phi(t,{\bm x})=\phi(t)+\delta\phi(t,{\bm x}),
\qquad
h_{ij}(t)=a(t)^2\delta_{ij}.
\end{equation}
We assume that vector and tensor perturbations are negligible at leading order, and study only the gravitational waves generated by scalar modes at second order after the end of inflation. We work with the following conventions for the slow-roll parameters,
\begin{equation}
\epsilon=-\frac{\dot{H}}{H^2},
\qquad
\eta=-\frac{1}{2}\frac{\dot{\epsilon}}{H\epsilon},
\end{equation}
where dots denote derivatives with respect to cosmic time $dt$.\footnote{Note that one must be careful when comparing different references, as another common convention is to define $\eta = \dot{\epsilon}/(H\epsilon)$.}. We are interested in a scenario in which the inflaton is initially slowly rolling down the potential until it reaches an inflection point at low field values, triggering the start of a USR phase in which the second slow-roll parameter $\eta$ becomes $\mathcal{O}(1)$, followed by a constant-roll (CR) phase in which $\eta$ switches sign and attains a constant negative value that it maintains until the end of inflation \cite{Kinney:2005vj,Motohashi:2014ppa}. The first slow-roll parameter $\epsilon$ remains small throughout the entire evolution. The terminology we adopt in this work is to denote the phase of $\eta>3/2$ following the initial SR stage by USR, and to denote the subsequent phase of $\eta<0$ by CR.

\begin{figure}
\centering
\includegraphics[width=0.49 \textwidth]{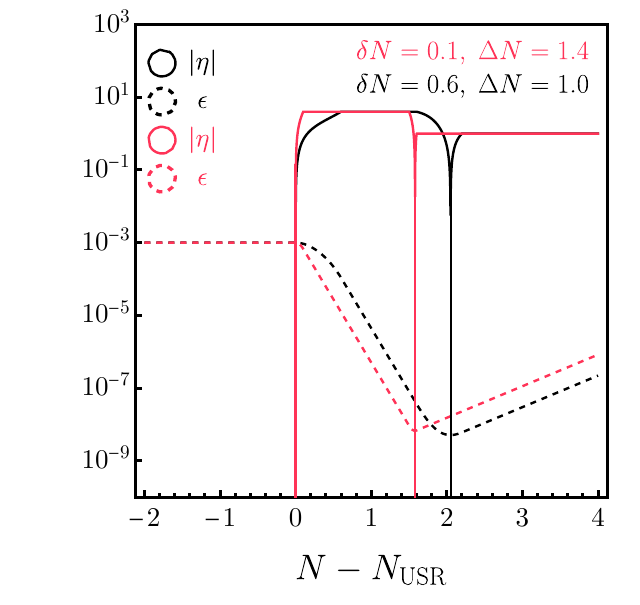} 
\caption{
\it Slow-roll parameters $\eta$ (solid) and $\epsilon$ (dashed) for the model considered in this paper (see Section \ref{sec:numerical}). The red lines correspond to $\delta N=0.1$ and $\Delta N=1.4$, and the black ones to $\delta N=0.6$ and $\Delta N=1.0$. The second slow-roll parameter vanishes in the initial SR phase, then becomes large and positive during the USR stage (which begins at $N_{\rm USR}$), and attains a constant negative value during the final CR phase.
}
\label{fig:pars_ref}
\end{figure}

The energy density of gravitational waves involves computing the two point function $\langle h^2\rangle$ of the tensor modes. Since tensor modes are sourced by terms quadratic in scalars, this involves obtaining the four-point function $\langle\delta\phi^4\rangle$, which in turn implies that we need the action up to fourth order in perturbations. The calculation of the full fourth-order action is a lengthy affair, but, as noted in \cite{Inomata:2022yte,Ballesteros:2024zdp}, it turns out that all of the terms resulting from the interaction between the inflaton and the gravitational sector are suppressed by powers of $\epsilon$ and can therefore be neglected.\footnote{To some extent, this is a model-dependent statement, which we take as an assumption. The assumption is satisfied for the model we describe in Section \ref{sec:numerical}.} The only terms one must keep track of are those involving derivatives of the inflaton potential, which contain time derivatives of $\eta$ and could be relevant during the USR phase. One can easily check that at $n$-th order the above action contains a trivial term $V_n\delta\phi^n/n!$ (where $V_n\equiv\partial_\phi^n V$) arising from the expansion of the potential, whereas the constraint equations for $N$ and $N^i$ involve only factors of $V_{n-2}$, since, due to the character of the lapse and shift as Lagrange multipliers, we always need to solve the constraint equations to two orders less than the one we are interested in, as first noted in \cite{Maldacena:2002vr}. Since $V_{n-2}/V_n\sim \epsilon$, in the limit $\epsilon\rightarrow 0$ the only surviving term is the one arising from the self-interaction of the inflaton fluctuations $V_n\delta\phi^n/n!$ \cite{Ballesteros:2024zdp}. Throughout the paper we therefore consider the following action
\begin{equation}
S
=
\int d^4x \bigg[
\frac{a^3}{2}\delta\dot{\phi}^2
-\frac{a}{2}(\partial\delta\phi)^2
-a^3\sum_n\frac{V_n}{n!}\delta\phi^n
\bigg].
\label{eq:action_deltaphi}
\end{equation}

The quantity we are interested in is the gauge-invariant curvature perturbation $\mathcal{R}$. One can work from the start in a gauge in which the inflaton fluctuations vanish and the curvature perturbation is obtained by perturbing the metric,
\begin{equation}
\delta\phi(t,{\bm x})=0,
\qquad
h_{ij}(t)=a(t)^2e^{2\mathcal{R}(t,{\bm x})}\delta_{ij}.
\end{equation}
This is the so-called $\mathcal{R}$ gauge, which comes with the disadvantage that the interactions relevant in USR are more difficult to identify. A better alternative is to work with the action (\ref{eq:action_deltaphi}) and perform a gauge transformation from the $\delta\phi$ gauge to the $\mathcal{R}$ gauge to obtain the curvature perturbation. Since we are interested in modes of $\mathcal{R}$ that are superhorizon at the end of inflation, we can neglect gradients in the transformation (which in turn implies that the spatial part of the coordinate transformation required to go from one gauge to the other is also irrelevant, see \cite{Ballesteros:2024zdp}). If we use a time reparameterization $\tilde{t}=t+\delta t$ and ask that the field transforms as
\begin{equation}
\Phi(t+\delta t)=\tilde{\Phi}(t)
\qquad\longrightarrow\qquad
\phi(t+\delta t)+\delta\phi(t+\delta t)=\phi(t),
\label{eq:field_trans}
\end{equation}
where $\tilde{\Phi}$ denotes the field in the $\mathcal{R}$ gauge, we can obtain $\delta t$ as a power series in $\delta\phi$. Similarly, on superhorizon scales $\mathcal{R}$ is space-independent, so by comparing both metrics we find
\begin{equation}
a(t)^2e^{2\mathcal{R}}
=
a(t+\delta t)^2
\qquad\longrightarrow\qquad
\mathcal{R}=\log\bigg[\frac{a(t+\delta t)}{a(t)}\bigg].
\end{equation}
Since we have $\delta t$ as a power series in $\delta\phi$ from eq.\,(\ref{eq:field_trans}), after expanding, this equation lets us write the curvature perturbation as
\begin{equation}
\mathcal{R}
=
\varphi
-\frac{1}{2}\eta\varphi^2
+\frac{1}{3}\bigg(\eta^2+\frac{\dot{\eta}}{2H}\bigg)\varphi^3
+\mathcal{O}(\varphi^4),
\end{equation}
where we have defined the linearized curvature perturbation
\begin{equation}
\varphi
\equiv
-\frac{H}{\dot{\phi}}\delta\phi.
\label{eq:linear_r}
\end{equation}

In fact, the entire series can be resummed in a straightforward way by noticing that, in the presence of a USR phase, modes evolve outside the horizon and only freeze to their final values after the phase is over, during the following CR period in which $\eta$ attains a constant value. It is then easy to show that the above series adds up to
\begin{equation}
\mathcal{R}=\frac{\log(1+\eta_{\rm CR}\varphi)}{\eta_{\rm CR}},
\label{eq:delta_n_formalism}
\end{equation}
where $\eta_{\rm CR}$ is the constant value of $\eta$ after the USR phase \cite{Atal:2019cdz,Ballesteros:2024pwn}. This compact formula is nothing but the standard result of the $\delta N$ formalism obtained from the point of view of perturbation theory.\footnote{This formula leads to an exponential tail in the PDF of $\mathcal{R}$ if $\varphi$ is assumed to be Gaussian \cite{Ballesteros:2024pwn}, a result that has also been obtained in the context of stochastic inflation \cite{Ezquiaga:2019ftu,Figueroa:2021zah}.} This equation can be expanded in a series of the form\footnote{These parameters are often defined with factors of $3/5$ in front for historical reasons. We refrain from using this convention, but one must be careful when comparing different references.}
\begin{equation}
\mathcal{R}=\varphi + f_{\rm NL}\varphi^2 + g_{\rm NL}\varphi^3+\cdots,
\label{eq:r_nonlin}
\end{equation}
where $f_{\rm NL}=-\eta_{\rm CR}/2$ and $g_{\rm NL}=\eta_{\rm CR}^2/3$. If the variable $\varphi$ is Gaussian, these are non-Gaussianities of the local form. The effect of these parameters on the induced GW spectrum for a Gaussian $\varphi$ has been studied extensively in the literature, see e.g.\,\cite{Garcia-Bellido:2017aan,Cai:2018dig,Unal:2018yaa,Bartolo:2019zvb,Atal:2021jyo,Adshead:2021hnm,Ragavendra:2021qdu,Abe:2022xur,Li:2023qua,Li:2023xtl,Papanikolaou:2024kjb}.\footnote{In \cite{Li:2023xtl} and other works, these nonlinearity parameters are often kept completely free, and large values are chosen for some of the numerical examples. We see, however, that in the presence of a USR phase they are not independent, but are related to each other through $\eta_{\rm CR}$, which is necessarily a negative $\mathcal{O}(1)$ number.} Throughout the rest of the paper we will discuss what happens if $\varphi$ is not a Gaussian variable, but instead has self-interaction terms given by eq.\,(\ref{eq:action_deltaphi}). The correlators of $\varphi$ can be obtained by using the in-in formalism (see e.g.\,\cite{Weinberg:2005vy,Ballesteros:2024zdp} for pedagogical reviews), and the correlators of $\mathcal{R}$ can be subsequently found by using eq.\,(\ref{eq:r_nonlin}). The expectation value of a Hermitian operator $Q$ is given by
\begin{equation}
\langle Q(t)\rangle
=
i^n\int_{-\infty}^tdt_1\cdots\int_{-\infty}^{t_{n-1}}dt_n
\langle
\big[H_I(t_n),\big[H_I(t_{n-1}),\cdots,\big[H_I(t_1),Q_I(t)\big]\cdots\big]\big]
\rangle,
\label{eq:in-in-master}
\end{equation}
where $H_I$ denotes the interaction Hamiltonian in the interaction picture, and $Q_I$ the interaction picture operator. This equation can also be rewritten as \cite{Ballesteros:2024zdp}
\begin{align*}
\langle Q(t)\rangle
=
\langle Q_I(t)\rangle
&+
2\,{\rm Im}\bigg\{\int_{-\infty}^t dt'
\langle Q_I(t)H_I(t'_-)\rangle\bigg\}\\
&+
2\,{\rm Re}\bigg\{\int_{-\infty}^t dt'\int_{t'}^t dt''
\langle \big[H_I(t_+'')Q_I(t)-Q_I(t)H_I(t_-'')\big]H_I(t'_-)\rangle\bigg\},
\numberthis
\end{align*}
where $\tau_\pm=\tau(1\mp i\omega)$ encapsulates the effect of the $i\omega$ prescription, necessary for the convergence of the integrals in the subhorizon regime (see the discussion in Section \ref{sec:numerical}).

\subsection{Loop corrections to the power spectrum}
\label{sec:one_loop_spectrum}

Let us address the renormalization of the Lagrangian (\ref{eq:action_deltaphi}) following closely the discussion in \cite{Ballesteros:2024zdp}. Assuming the Lagrangian contains only bare fields and couplings, these can be renormalized by defining the appropriate counterterms, $\delta\phi_b=Z_\phi^{1/2}\delta\phi_r$ and $V_{nb}=Z_{V_n}Z_\phi^{-n/2}V_{nr}$. We can then expand
\begin{equation}
Z_\phi=1+\delta_\phi,\qquad Z_{V_n}=1+\delta_{V_n}.
\end{equation}
Keeping all the terms that affect the two-point function of $\delta\phi$ at one loop, we find the following interaction Hamiltonian
\begin{equation}
H_{\rm I}
=
\int d^3x \bigg[
a^3\bigg(\frac{V_3}{3!}\delta\phi^3+\frac{V_4}{4!}\delta\phi^4\bigg)
+\frac{a^3}{2}\delta_\phi\delta\dot{\phi}^2
+\frac{a}{2}\delta\phi
\Big(
a^2\tilde{\delta}_V
-\delta_\phi\partial^2
\Big)
\delta\phi
\bigg],
\end{equation}
where $\tilde{\delta}_V\equiv \delta_{V_2}V_2$. Instead of working with the potential derivatives, we find it convenient to define
\begin{equation}
v_n\equiv (2\epsilon)^{n/2}V_n\big|_{\epsilon\rightarrow 0},
\end{equation}
where $v_n/H^2$ is, in general, some dimensionless combination of slow-roll parameters, and we have taken the $\epsilon\rightarrow 0$ limit to be consistent with our approximation of neglecting all $\epsilon$-suppressed interactions. Thus,
\begin{align}
v_3&=2\epsilon\bigg[
(3-2\eta)\frac{d\eta}{dN}+\frac{d^2\eta}{dN^2}
\bigg]H^2,
\\
v_4&=2\epsilon\bigg[
\frac{d^3\eta}{dN^3}+(3-\eta)\frac{d^2\eta}{dN^2}
-2\bigg(\frac{d\eta}{dN}\bigg)^2+(3-2\eta)\eta\frac{d\eta}{dN}
\bigg]H^2,
\end{align}
where the derivatives are taken with respect to the number of $e$-folds $dN=Hdt$.

We are now ready to introduce the first set of Feynman rules we use throughout this work, following an approach similar to that of \cite{Ballesteros:2024zdp,Li:2023xtl}. It is more convenient to work with the linearized curvature perturbation $\varphi$ defined in eq.\,(\ref{eq:linear_r}) than with $\delta\phi$. Fields are represented by
\begin{equation}
\mathcal{R}
\quad\sim\quad
\begin{tikzpicture}[baseline={-2}]
    \draw [dashed] (0,0) -- (1.5,0);
\end{tikzpicture}\;,
\hspace{0.5in}
\varphi
\quad\sim\quad
\begin{tikzpicture}[baseline={-2}]
    \draw (0,0) -- (1.5,0);
\end{tikzpicture}\;.
\end{equation}
The self-interactions of the inflaton are represented by white dots,
\begin{equation}
v_3
\quad\sim\quad
\begin{tikzpicture}[baseline={-2}]
    \draw (0,0) -- (0.56,0);
    \draw (0.56,0) -- (0.81,0.5);
    \draw (0.56,0) -- (0.81,-0.5);
	\fill[black] (0.56,0) circle (2.5pt);
	\fill[white] (0.56,0) circle (1.5pt);
\end{tikzpicture}\;,
\hspace{0.5in}
v_4
\quad\sim\quad
\begin{tikzpicture}[baseline={-2}]
    \draw (0,0) -- (0.56,0);
    \draw (0.56,0) -- (0.81,0.5);
    \draw (0.56,0) -- (0.81,-0.5);
    \draw (0.56,0) -- (1.12,0);
	\fill[black] (0.56,0) circle (2.5pt);
	\fill[white] (0.56,0) circle (1.5pt);
\end{tikzpicture}\;,
\end{equation}
and the interactions arising from the change of gauge by black dots,
\begin{equation}
1
\quad\sim\quad
\begin{tikzpicture}[baseline={-2}]
    \draw [dashed] (0,0) -- (0.7,0);
    \draw (0.7,0) -- (1.4,0);
	\fill[black] (0.71,0) circle (2pt);
\end{tikzpicture}\;,
\hspace{0.5in}
f_{\rm NL}
\quad\sim\quad
\begin{tikzpicture}[baseline={-2}]
    \draw [dashed] (0,0) -- (0.56,0);
    \draw (0.56,0) -- (0.81,0.5);
    \draw (0.56,0) -- (0.81,-0.5);
	\fill[black] (0.56,0) circle (2pt);
\end{tikzpicture}\;,
\hspace{0.5in}
g_{\rm NL}
\quad\sim\quad
\begin{tikzpicture}[baseline={-2}]
    \draw [dashed] (0,0) -- (0.56,0);
    \draw (0.56,0) -- (0.81,0.5);
    \draw (0.56,0) -- (0.81,-0.5);
    \draw (0.56,0) -- (1.12,0);
	\fill[black] (0.56,0) circle (2pt);
\end{tikzpicture}\;.
\label{eq:gauge_ints}
\end{equation}
Solid lines connected to two black dots on either side correspond to propagators,\begin{equation}
|\varphi_k|^2
\quad\sim\quad
\begin{tikzpicture}[baseline={-2}]
	\fill[black] (0,0) circle (2pt);
    \draw (0,0) -- (1.5,0);
	\fill[black] (1.5,0) circle (2pt);
\end{tikzpicture}\;,
\end{equation}
whereas lines connected to white dots have a more complicated structure and require explicitly expanding eq.\,(\ref{eq:in-in-master}). We have the following rule for one white dot connected to three black dots
\begin{equation}
-\int_{-\infty}^t dt' a(t')^3v_3(t')
\,2{\rm Im}
\Big[
\varphi_p(t)\varphi^\star_p(t')
\varphi_q(t)\varphi^\star_q(t')
\varphi_k(t)\varphi^\star_k(t')
\Big]
\quad\sim\quad
\begin{tikzpicture}[baseline={-2}]
    \draw (0,0) -- (0.56,0);
    \draw (0.56,0) -- (0.81,0.5);
    \draw (0.56,0) -- (0.81,-0.5);
	\fill[black] (0.56,0) circle (2.5pt);
	\fill[white] (0.56,0) circle (1.5pt);
	\fill[black] (0,0) circle (2pt);
	\fill[black] (0.81,0.5) circle (2pt);
	\fill[black] (0.81,-0.5) circle (2pt);
\end{tikzpicture}\;,
\end{equation}
and the following rule for two white dots connected to four black dots
\begin{align*}
\int_{-\infty}^t dt' a(t')^3v_3(t')\int_{t'}^t a(t'')^3v_3(&t'')dt''
\\
\Big\{2{\rm Im}\Big[
\varphi_k(t)
\varphi^\star_k(t'')
\varphi_\ell(t)
\varphi^\star_\ell(t'')
\Big]
&2{\rm Im}\Big[
\varphi_p(t)
\varphi^\star_p(t')
\varphi_q(t)
\varphi^\star_q(t')
\varphi_{|{\bm p}+{\bm q}|}(t'')
\varphi^\star_{|{\bm p}+{\bm q}|}(t')
\Big]
\\
+2{\rm Im}\Big[
\varphi_p(t)
\varphi^\star_p(t'')
\varphi_q(t)
\varphi^\star_q(t'')
\Big]
&2{\rm Im}\Big[
\varphi_k(t)
\varphi^\star_k(t')
\varphi_\ell(t)
\varphi^\star_\ell(t')
\varphi_{|{\bm p}+{\bm q}|}(t'')
\varphi^\star_{|{\bm p}+{\bm q}|}(t')
\Big]\Big\}
\quad\sim\quad
\begin{tikzpicture}[baseline={-2}]
	\fill[black] (0,+0.5) circle (2pt);
	\fill[black] (1,+0.5) circle (2pt);
	\fill[black] (0,-0.5) circle (2pt);
	\fill[black] (1,-0.5) circle (2pt);
    \draw (0,0.5) -- (0.5,0.25);
    \draw (0.5,0.25) -- (1,0.5);
    \draw (0,-0.5) -- (0.5,-0.25);
    \draw (0.5,-0.25) -- (1,-0.5);
    \draw (0.5,0.25) -- (0.5,-0.25);
	\fill[black] (0.5,0.25) circle (2.5pt);
	\fill[white] (0.5,0.25) circle (1.5pt);
	\fill[black] (0.5,-0.25) circle (2.5pt);
	\fill[white] (0.5,-0.25) circle (1.5pt);
\end{tikzpicture}\;,
\numberthis
\end{align*}
where the momentum $|{\bm p}+{\bm q}|$ corresponds to the internal line and changes depending on whether one considers the $s$, $t$ or $u$ channel.

In these expressions and from now on, whenever we write the Fourier-space field $\varphi_k$ we refer to the modes of the free field, obtained by solving the Mukhanov-Sasaki equation
\begin{equation}
\ddot{\varphi}_k+(3-2\eta)H\dot{\varphi}_k+\frac{k^2}{a^2}\varphi_k=0
\end{equation}
with Bunch-Davies initial conditions. Finally, the quadratic counterterm is represented by
\begin{equation}
\delta_\varphi
\quad\sim\quad
\begin{tikzpicture}[baseline={-2}]
    \draw (0,0) -- (1.4,0);
	\fill[black] (0.7,0) circle (5pt);
	\fill[white] (0.7,0) circle (4.25pt);
	\node[black] at (0.7,0) {$\times$};
\end{tikzpicture}\;.
\end{equation}
An explicit expression for $\delta_\varphi$ in terms of $\delta_\phi$ and $\tilde{\delta}_V$ will be given below.

Let us discuss the renormalization of the one-loop propagator (i.e.\,the power spectrum) $\langle\hat{\mathcal{R}}^2\rangle$ following the procedure of \cite{Ballesteros:2024zdp}. Seven diagrams contribute to the two-point function at one loop. The first is the tree-level graph
\begin{equation}
\frac{2\pi^2}{k^3}
\mathcal{P}_\mathcal{R}(k)
\quad=\quad
\begin{tikzpicture}[baseline={-2}]
    \draw [dashed] (0,0) -- (0.5,0);
    \draw [dashed] (2,0) -- (2.5,0);
    \draw (0.5,0) -- (2,0);
	\fill[black] (2,0) circle (2pt);
    \fill[black] (0.5,0) circle (2pt);
\end{tikzpicture}
\quad=\quad
|\varphi_k(t)|^2.
\end{equation}
Note that for each external dashed line corresponding to $\mathcal{R}$ we must first decide how to turn it into a solid line $\varphi$ using (\ref{eq:gauge_ints}), and then connect the lines using the interaction vertices (in a similar manner to \cite{Li:2023xtl}). There are two one-loop diagrams involving only gauge-transformation vertices, which yield the following corrections
\begin{align}
\begin{tikzpicture}[baseline={-2}]
    \draw [dashed] (0,0) -- (0.9,0);
    \draw [dashed] (1.6,0) -- (2.5,0);
    \draw (1.25,0) circle (10pt);
	\fill[black] (0.9,0) circle (2pt);
	\fill[black] (1.6,0) circle (2pt);
\end{tikzpicture}
\quad &= \quad
2f_{\rm NL}^2
\int\frac{d^3p}{(2\pi)^3}
|\varphi_{|{\bm k}-{\bm p}|}(t)|^2
|\varphi_{p}(t)|^2,
\label{eq:fg_loops_1}
\\
\begin{tikzpicture}[baseline={-2}]
    \draw [dashed] (0,0) -- (0.9,0);
    \draw [dashed] (1.6,0) -- (2.5,0);
    \draw (0.9,0) -- (1.6,0);
    \draw (0.9,0.35) circle (10pt);
	\fill[black] (0.9,0) circle (2pt);
	\fill[black] (1.6,0) circle (2pt);
\end{tikzpicture}
\quad &= \quad
6g_{\rm NL}|\varphi_k(t)|^2\int\frac{d^3p}{(2\pi)^3}|\varphi_p(t)|^2,
\label{eq:fg_loops_2}
\end{align}
where we have also accounted for the corresponding symmetry factors. There are three diagrams with self-interactions which involve one time integral for each white vertex, as per the in-in formula in eq.\,(\ref{eq:in-in-master}),
\begin{align*}
\begin{tikzpicture}[baseline={-2}]
    \draw [dashed] (0,0) -- (0.55,0);
    \draw [dashed] (2,0) -- (2.5,0);
    \draw (1.25,0) -- (2,0);
    \draw (0.9,0) circle (10pt);
	\fill[black] (2,0) circle (2pt);
    \fill[black] (0.55,0) circle (2pt);
	\fill[black] (1.25,0) circle (2.5pt);
	\fill[white] (1.25,0) circle (1.5pt);
\end{tikzpicture}
\quad &= \quad
-2f_{\rm NL}
\int_{-\infty}^t a(t')^3v_3(t') dt'
\\&
\hspace{1in}
\int\frac{d^3p}{(2\pi)^3}
2{\rm Im}\Big[
\varphi_{|{\bm k}-{\bm p}|}(t)
\varphi^\star_{|{\bm k}-{\bm p}|}(t')
\varphi_k(t)
\varphi^\star_k(t')
\varphi_p(t)
\varphi^\star_p(t')
\Big],\numberthis
\label{eq:cubic_loop_1}
\\
\begin{tikzpicture}[baseline={-2}]
    \draw [dashed] (0,0) -- (0.5,0);
    \draw (0.5,0) -- (0.9,0);
    \draw [dashed] (2,0) -- (2.5,0);
    \draw (1.6,0) -- (2,0);
    \draw (1.25,0) circle (10pt);
	\fill[black] (0.5,0) circle (2pt);
	\fill[black] (2,0) circle (2pt);
    \fill[black] (0.9,0) circle (2.5pt);
	\fill[white] (0.9,0) circle (1.5pt);
	\fill[black] (1.6,0) circle (2.5pt);
	\fill[white] (1.6,0) circle (1.5pt);
\end{tikzpicture}
\quad &= \quad
\int_{-\infty}^t a(t')^3v_3(t') dt'
\int_{t'}^{t} a(t'')^3v_3(t'') dt''
2{\rm Im}\Big[
\varphi_k(t)
\varphi^\star_k(t'')
\Big]
\\&\hspace{1in}
\int\frac{d^3p}{(2\pi)^3}
2{\rm Im}\Big[
\varphi_k(t)\varphi^\star_k(t')
\varphi_{|{\bm k}-{\bm p}|}(t'')\varphi^\star_{|{\bm k}-{\bm p}|}(t')
\varphi_p(t'')\varphi^\star_p(t')
\Big],\numberthis
\label{eq:cubic_loop_2}
\\
\begin{tikzpicture}[baseline={-2}]
    \draw [dashed] (0,0) -- (0.5,0);
    \draw [dashed] (2,0) -- (2.5,0);
    \draw (0.5,0) -- (2,0);
    \draw (1.25,0.35) circle (10pt);
	\fill[black] (1.25,0) circle (2.5pt);
	\fill[white] (1.25,0) circle (1.5pt);
	\fill[black] (0.5,0) circle (2pt);
	\fill[black] (2,0) circle (2pt);
\end{tikzpicture}
\quad &= \quad
\frac{1}{2}\int_{-\infty}^t 
a(t')^3v_4(t') dt'
2{\rm Im}\Big\{\Big[
\varphi_k(t)
\varphi^\star_k(t')
\Big]^2\Big\}
\int \frac{d^3p}{(2\pi)^3}|\varphi_p(t')|^2.\numberthis
\label{eq:quartic_loop}
\end{align*}
The negative sign in the first diagram arises from switching between $\delta\phi$ and $\varphi$ in the interaction Hamiltonian. Finally, the diagram corresponding to the counterterm is
\begin{equation}
\begin{tikzpicture}[baseline={-2}]
    \draw [dashed] (0,0) -- (0.5,0);
    \draw [dashed] (2,0) -- (2.5,0);
    \draw (0.5,0) -- (2,0);
	\fill[black] (2,0) circle (2pt);
    \fill[black] (0.5,0) circle (2pt);
	\fill[black] (1.25,0) circle (5pt);
	\fill[white] (1.25,0) circle (4.25pt);
	\node[black] at (1.25,0) {$\times$};
\end{tikzpicture}
\quad = \quad
(\propto\delta_\phi)
+
\int_{-\infty}^t dt'
a\epsilon
(a^2\tilde{\delta}_V+\delta_\phi k^2)
2{\rm Im}\Big\{\Big[
\varphi_k(t)\varphi^\star_k(t')
\Big]^2\Big\},
\label{eq:counterterm}
\end{equation}
where the first term is proportional to $\delta_\phi$ and its exact form is irrelevant in what follows. Due to general covariance, none of the $\delta_i$ terms can depend on the spatial coordinates ${\bm x}$, but they are allowed to depend on $t$ due to the fact that the background is time-dependent. Moreover, due to the fact that we do not work with a specific potential but rather write it as an arbitrary function of time through the slow-roll parameters, the $\delta_{V_n}$ terms have a completely arbitrary time dependence \cite{Ballesteros:2024zdp}. Due to this arbitrary time dependence, the divergence in the diagram involving a quartic interaction can be completely absorbed by the counterterm.\footnote{In other words, since we do not have an explicit functional form for the potential in terms of coupling constants, but rather an arbitrary function of time reconstructed from the slow-roll parameters, the counterterms have an arbitrary time dependence that allows us to absorb the divergence (our parameterization for the potential is, in a rough sense, equivalent to having coupling constants with an arbitrary time dependence in the Lagrangian).} Comparing eqs.\,(\ref{eq:quartic_loop}) and (\ref{eq:counterterm}), we have $\delta_\phi=0$ and\footnote{Note that this is essentially equivalent to using a minimal subtraction scheme to eliminate the divergence. The finite part of the diagram must be fixed in the usual manner by imposing renormalization conditions. In the spirit of \cite{Ballesteros:2024zdp}, we do not go through with this procedure and rather assume that this contribution is at most of the same order as the remaining finite diagrams.}
\begin{equation}
\tilde{\delta}_V(t)
=
\frac{v_4(t)}{2\epsilon(t)}
\int \frac{d^3p}{(2\pi)^3}|\varphi_p(t)|^2.
\end{equation}
We remark that this is only possible because the right-hand side of the above equation does not depend on the external momentum $k$. The same cannot be done for the first two diagrams with self-interactions, which would require the counterterm to be $k$-dependent. Fortunately, these two diagrams turn out to be completely finite, and therefore yield a nonvanishing correction to the curvature power spectrum independent of renormalization conditions.

\section{Induced gravitational waves}
\label{sec:waves}

In this Section we calculate the non-Gaussian corrections to the spectrum of induced gravitational waves in the presence of a USR phase. We first review the standard calculation of the Gaussian piece of the gravitational wave energy density, and introduce the corresponding Feynman rules. We then proceed to determine all the relevant two-loop diagrams and write down the corresponding integrals.

\subsection{Gravitational wave energy density}

At leading order in perturbations, scalar and tensor modes evolve independently. Once Einstein's equations are expanded to second order, however, we find that gravitational waves are sourced by terms quadratic in scalar modes.\footnote{See \cite{Espinosa:2018eve,Domenech:2021ztg} for pedagogical reviews of the calculation.} In Fourier space, the transverse-traceless tensor modes of the metric can be written as
\begin{equation}
h_{ij}({\bm x})
=
\int\frac{d^3k}{(2\pi)^3}
e^s_{ij}({\bm k})h_{\bm k}^s
e^{i{\bm k}\cdot{\bm x}},
\label{eq:tensor_fourier}
\end{equation}
where the index $s=+,\times$ refers to the two tensor polarizations, and the quantities $e^s_{ij}({\bm k})$ are symmetric, transverse-traceless tensors defined via
\begin{align}
\label{eq:pol_tensors}
e^+_{ij}=\frac{1}{\sqrt{2}}(v_iv_j-\bar{v}_i\bar{v}_j),\quad		
e^\times_{ij}=\frac{1}{\sqrt{2}}(v_i\bar{v}_j+\bar{v}_iv_j),
\end{align}
where $\bm{v}$ and $\bar{\bm{v}}$ are two unit vectors satisfying $\bm{k}\cdot\bm{v}=\bm{k}\cdot\bar{\bm{v}}=\bm{v}\cdot\bar{\bm{v}}=0$. 

The equation of motion for $h_k^s$ is
\begin{equation}
h_k^{s\prime\prime}+2\mathcal{H}h_k^{s\prime}+k^2h_k^s=S_k^s,
\label{eq:tensor_eom}
\end{equation}
where $\mathcal{H}=a'/a$, primes denote derivatives with respect to conformal time $d\tau=dt/a$, and the source term $S_k^s$ is, in the Newtonian gauge and in the absence of anisotropic stress,
\begin{equation}
S^s_k=\int\frac{d^3p}{(2\pi)^3}
\Big[{\bm p}\cdot {\bm e}^s({\bm k})\cdot{\bm p}\Big]
\bigg[8\psi_{\bm p}\psi_{{\bm k}-{\bm p}}+\frac{16}{3(1+w)}\bigg(\psi_{\bm p}+\frac{\psi'_{\bm p}}{\mathcal{H}}\bigg)\bigg(\psi_{{\bm k}-{\bm p}}+\frac{\psi'_{{\bm k}-{\bm p}}}{\mathcal{H}}\bigg)\bigg],
\label{eq:source}
\end{equation}
with $w=p/\rho$ being the equation of state of the fluid that dominates the Universe and $\psi$ representing the Newtonian potential.

In this work we focus on gravitational waves induced after the end of inflation, once the perturbations re-enter the horizon. Assuming that a negligible amount of gravitational waves has been induced by the time inflation ends,\footnote{
The gravitational waves induced during inflation have been considered e.g.\,in \cite{Fumagalli:2021mpc,Firouzjahi:2023btw,Ballesteros:2024qqx}. Due to the appearance of additional UV divergences in that case because of the Bunch-Davies initial conditions, the calculation of the corresponding loop corrections becomes more involved and necessitates a more careful treatment, so we do not consider it here. We point the reader to \cite{Ballesteros:2024qqx} for recent work on the subject. We nevertheless remark that the corresponding contribution of the inflationary term to the spectrum can be easily included as an additional term in $\Omega_{\rm GW}$, see e.g.\,\cite{Ballesteros:2022hjk}.} we can write the solution to the equation of motion as
\begin{equation}
h_k^s(\tau)=\int_0^\tau G_k(\tau,\tau')S_k^s(\tau')d\tau',
\end{equation}
where $\tau=0$ corresponds to the end of inflation. This solution can also be written as
\begin{equation}
h_k^s(\tau)
=
\frac{1}{k^2}\int\frac{d^3p}{(2\pi)^3}
\Big[{\bm p}\cdot {\bm e}^s({\bm k})\cdot{\bm p}\Big]
\mathcal{R}_{\bm p}\mathcal{R}_{{\bm k}-{\bm p}}
I_k(\tau,p,|{\bm k}-{\bm p}|),
\label{eq:h_solution}
\end{equation}
where ${\bm p}\cdot{\bm e}^s({\bm k})\cdot{\bm p}=e^s_{ij}(k)p_ip_j$ and we have defined
\begin{equation}
I_k(\tau,p,|{\bm k}-{\bm p}|)
=
\bigg(\frac{3+3w}{5+3w}\bigg)^2
\int kG_k(\tau,\tau')Q_k(\tau',p,|{\bm k}-{\bm p}|) kd\tau',
\end{equation}
with
\begin{align*}
Q_k(\tau,p,|{\bm k}-{\bm p}|)
&=
8T_{\psi}(p\eta)T_{\psi}(|{\bm k}-{\bm p}|\tau)
\\&
\hspace{0.5in}
+\frac{16}{3(1+w)}
\bigg[T_{\psi}(p\tau)+\frac{T'_{\psi}(p\tau)}{\mathcal{H}}\bigg]\bigg[T_{\psi}(|{\bm k}-{\bm p}|\tau)+\frac{T'_{\psi}(|{\bm k}-{\bm p}|\tau)}{\mathcal{H}}\bigg].
\numberthis
\end{align*}
In these expressions we have defined the scalar transfer function as $\psi_k(\tau)=T_\psi(k\tau)\psi_k(0)$ and used the following relation between the initial value of the Newtonian potential and the frozen curvature perturbation on superhorizon scales
\begin{equation}
\label{eq:newton_initial}
\psi_{\bm k}(0)=\frac{3+3w}{5+3w}\mathcal{R}_{\bm k}.
\end{equation}

The Green's function can be obtained by finding two linearly independent solutions $h_1$ and $h_2$ to the homogeneous equation (\ref{eq:tensor_eom}) and taking
\begin{equation}
G_k(\tau,\tau')
=
\frac{h_1(\tau)h_2(\tau')-h_1(\tau')h_2(\tau)}{h'_1(\tau')h_2(\tau')-h_1(\tau')h'_2(\tau')}.
\end{equation}
During radiation domination $\mathcal{H}=1/\tau$ and two suitable solutions are
\begin{equation}
h_1(k\tau)=\frac{\sin(k\tau)}{k\tau}\qquad{\rm and}\qquad h_2(k\tau)=\frac{\cos(k\tau)}{k\tau}.
\end{equation}
The scalar transfer function, on the other hand, is obtained by solving 
\begin{equation}
\psi''_k+3(1+w)\mathcal{H}\psi'_k+wk^2\psi_k=0
\end{equation}
with boundary conditions $\psi_k(0)=1$ and $\psi'_k(0)=0$. During radiation domination we have
\begin{equation}
T_\psi(k\tau)
=
\bigg(\frac{3}{k\tau}\bigg)^2\bigg[
\frac{\sqrt{3}}{k\tau}\sin\bigg(\frac{k\tau}{\sqrt{3}}\bigg)
-\cos\bigg(\frac{k\tau}{\sqrt{3}}\bigg)
\bigg].
\end{equation}

If we quantize $h_{\bm k}^s\rightarrow \hat{h}_{\bm k}^s$, we can define the tensor power spectrum as
\begin{equation}
\langle \hat{h}_{\bm k}^s\hat{h}_{\bm p}^t\rangle
=
\frac{2\pi^2}{k^3}\mathcal{P}_h(k)\delta^{st}(2\pi)^3\delta^3({\bm k}+{\bm p}).
\label{eq:tensor_spectrum}
\end{equation}
We now introduce the Feynman rules for the tensor modes. The field $h$ is represented by a wavy line, and the interaction in eq.\,(\ref{eq:h_solution}) can be thought of as a wavy line splitting into two dashed lines,\footnote{We could also introduce self-interactions for the tensor modes, but we neglect these due to the suppression of the leading-order tensor power spectrum with respect to the scalar one.}
\begin{equation}
h
\quad\sim\quad
\begin{tikzpicture}[baseline={-2}]
    \draw [decorate,decoration={snake,amplitude=1.5pt,segment length=6pt}] (0,0) -- (1.5,0);
\end{tikzpicture}\;,
\hspace{0.5in}
\frac{1}{k^2}
\Big[{\bm p}\cdot {\bm e}^s({\bm k})\cdot{\bm p}\Big]
I_k(p,|{\bm k}-{\bm p}|)
\quad\sim\quad
\begin{tikzpicture}[baseline={-2}]
    \draw [decorate,decoration={snake,amplitude=1.5pt,segment length=6pt}] (0,0) -- (0.7,0);
    \draw [dashed] (0.7,0) -- (1.2,0.5);
    \draw [dashed] (0.7,0) -- (1.2,-0.5);
\end{tikzpicture}\;.
\end{equation}
where here and from now on we omit the time dependence of $I_k$. At leading order (that is, at the Gaussian level), there is only one diagram contributing to the power spectrum for induced tensor modes,
\begin{equation}
2\frac{2\pi^2}{k^3}\mathcal{P}_h(k)
\quad=\quad
\begin{tikzpicture}[baseline={-2}]
    \draw [decorate,decoration={snake,amplitude=1.5pt,segment length=6pt}] (0,0) -- (0.7,0);
    \draw [dashed] (0.7,0) -- (1.2,0.5);
    \draw [dashed] (0.7,0) -- (1.2,-0.5);
	\fill[black] (1.2,0.5) circle (2pt);
	\fill[black] (2.2,0.5) circle (2pt);
	\fill[black] (1.2,-0.5) circle (2pt);
	\fill[black] (2.2,-0.5) circle (2pt);
    \draw (1.2,0.5) -- (2.2,0.5);
    \draw (1.2,-0.5) -- (2.2,-0.5);
    \draw [dashed] (2.2,0.5) -- (2.7,0);
    \draw [dashed] (2.2,-0.5) -- (2.7,0);
    \draw [decorate,decoration={snake,amplitude=1.5pt,segment length=6pt}] (2.7,0) -- (3.4,0);
\end{tikzpicture}\;,
\label{eq:one_loop_tensor}
\end{equation}
where the additional factor of $2$ comes from the sum over polarizations. This is nothing but the standard result for the induced gravitational wave spectrum,
\begin{equation}
2\frac{2\pi^2}{k^3}\mathcal{P}_h(k)
=
\frac{2}{k^4}
\int\frac{d^3p}{(2\pi)^3}
\Big[{\bm p}\cdot {\bm e}^s({\bm k})\cdot{\bm p}\Big]^2
I_k(p,|{\bm k}-{\bm p}|)^2|\varphi_p(t_e)|^2|\varphi_{|{\bm k}-{\bm p}|}(t_e)|^2,
\label{eq:power_h_tree}
\end{equation}
where the time $t_e$ corresponds to the end of inflation (or any arbitrary time at which the curvature perturbation is frozen on superhorizon scales). To perform the integral we fix ${\bm k}$ as the $\hat{\bm z}$ axis and use
\begin{equation}
{\bm p}\cdot {\bm e}^s({\bm k})\cdot{\bm p}
=
\frac{p^2}{\sqrt{2}}\sin^2\theta_p
\cdot
\begin{dcases}
    \cos 2\phi_p & {\rm if}\quad s=+, \\
    \sin 2\phi_p & {\rm if}\quad s=\times,
\end{dcases}
\end{equation}
where $\theta_p$ denotes the polar angle between ${\bm p}$ and ${\bm k}$, and $\phi_p$ the azimuthal angle.

Before moving on to the non-Gaussian corrections to this result, let us state the relation between the tensor power spectrum and the quantity of interest, the energy density of gravitational waves,
\begin{equation}
\label{eq:gw_energy}
\rho_{\rm GW}(\tau,{\bm x})=\frac{M_p^2}{16a^2}\overline{\partial_k h_{ij}({\bm x})\partial_k h_{ij}({\bm x})},
\end{equation}
where the bar denotes an average over time,\footnote{To be clear, this average applies only to the $I_k$ functions which carry the information about the time-evolution of the scalar perturbations inside the horizon. It does not apply to the initial conditions $\varphi_k$, which are frozen outside the horizon and should therefore be evaluated e.g.\,at the end of inflation.} which must be taken due to the stochastic nature of the signal \cite{maggiore}. In Fourier space, this becomes
\begin{equation}
\rho_{\rm GW}({\bm x})
=
\int\frac{d^3p}{(2\pi)^3}e^{i{\bm p}\cdot{\bm x}}
\frac{M_p^2}{16a^2}
\int\frac{d^3q}{(2\pi)^3}
(q^2-{\bm p}\cdot{\bm q})
\,{\rm Tr}
\Big[
{\bm e}^s({\bm p}-{\bm q})
\cdot
{\bm e}^t({\bm q})
\Big]
\overline{
h_{{\bm p}-{\bm q}}^s
h_{\bm q}^t
}.
\label{eq:gw_energy_fourier}
\end{equation}
We can obtain the total gravitational wave abundance by taking the ensemble average of this quantity and using eq.\,(\ref{eq:tensor_spectrum}),
\begin{equation}
\Omega_{\rm GW}(\tau)
\equiv
\frac{\langle\rho_{\rm GW}(\tau,{\bm x})\rangle}{\rho_c(\tau)}
=
\int d\log q
\bigg[\frac{1}{24}
\frac{q^2}{\mathcal{H}^2}
\overline{\mathcal{P}_h(q)}
\bigg]
\equiv
\int d\log q \,\Omega_{\rm GW}(q,\tau),
\end{equation}
where we have also used $e^s_{ij}({\bm k})e^s_{ij}({\bm k})=2$ and $\rho_c\equiv 3M_p^2H^2$. The time average of $I_k^2$ is, at late times,
\begin{equation}
\frac{k^2}{\mathcal{H}^2}\overline{I_k(p,|{\bm k}-{\bm p}|)^2}
=
\frac{1}{2}\Big[J_c(p,|{\bm k}-{\bm p}|)^2+J_s(p,|{\bm k}-{\bm p}|)^2\Big],
\end{equation}
where the right-hand side is time-independent and given by \cite{Espinosa:2018eve}
\begin{align}
J_c(p,|{\bm k}-{\bm p}|)
&=
\bigg(\frac{3+3w}{5+3w}\bigg)^2
36\pi\Theta_{\rm H}(s-1)\frac{(2-s^2-d^2)^2}{(s^2-d^2)^3},
\\
J_s(p,|{\bm k}-{\bm p}|)
&=
-\bigg(\frac{3+3w}{5+3w}\bigg)^2
36\frac{s^2+d^2-2}{(s^2-d^2)^2}\bigg[
\frac{s^2+d^2-2}{s^2-d^2}\log\bigg(\frac{1-d^2}{|s^2-1|}\bigg)+2
\bigg],
\end{align}
where, after setting $w=1/3$, we have used the shorthand
\begin{equation}
d
=
\frac{|p-|{\bm k}-{\bm p}||}{k\sqrt{3}},
\qquad
s
=
\frac{p+|{\bm k}-{\bm p}|}{k\sqrt{3}}.
\end{equation}

If gravitational waves are no longer generated towards the end of the radiation era, we can relate their abundance at that time to its present value by using entropy conservation,
\begin{equation}
\Omega_{\rm GW}(q,\tau)
=
\frac{\Omega_\gamma(T_0)}{24}\frac{g_\star(T)}{g_\star(T_0)}\bigg[\frac{g_{\star s}(T_0)}{g_{\star s}(T)}\bigg]^{4/3}
\frac{q^2}{\mathcal{H}^2}\overline{\mathcal{P}_h(q)},
\end{equation}
where $g_\star$ and $g_{\star s}$ denote the effective relativistic temperature and entropy degrees of freedom, respectively, $T_0$ is the temperature of radiation today, and $\Omega_\gamma(T_0)$ is the present radiation abundance. The temperature $T$ is some arbitrary time after the gravitational wave production stops, but before the top quark decouples from the primordial plasma.\footnote{We take $g_\star(T) = g_{\star s}(T) = 106.75$, $g_\star(T_0) = 3.36$, $g_{\star s}(T_0) = 3.94$, and $\Omega_\gamma(T_0)=5.8\cdot 10^{-5}$ \cite{Baumann_2022}.}

\subsection{Propagator at two loops}

If we assume that the first-order (free) tensor modes are negligible (see the discussion at the end of this Section), the induced gravitational wave spectrum at leading (Gaussian) order involves only the one-loop diagram in eq.\,(\ref{eq:one_loop_tensor}). Non-Gaussianities can be included by considering all possible connected two-loop diagrams\footnote{Our counting is different to that of \cite{Li:2023xtl}, where the series in eq.\,(\ref{eq:r_nonlin}) was truncated at order $g_{\rm NL}$ and all diagrams up to that order were considered. We instead organize the corrections to the propagator by the number of loops involved so that we can compare the relative strength of the different interactions.} of the form
\begin{equation}
2\frac{2\pi^2}{k^3}\mathcal{P}_h(k)
\quad = \quad
\begin{tikzpicture}[baseline={-2}]
    \draw [decorate,decoration={snake,amplitude=1.5pt,segment length=6pt}] (0,0) -- (0.7,0);
    \draw [dashed] (0.7,0) -- (1.2,0.5);
    \draw [dashed] (0.7,0) -- (1.2,-0.5);
    \draw[color=black] (1.2,-0.5) rectangle (2.2,0.5);
    \fill[pattern=north east lines, pattern color=black] (1.2,-0.5) rectangle (2.2,0.5);
	\fill[white] (1.2,+0.5) circle (3pt);
	\fill[black] (1.2,+0.5) circle (2pt);
 	\fill[white] (2.2,+0.5) circle (3pt);
	\fill[black] (2.2,+0.5) circle (2pt);
  	\fill[white] (1.2,-0.5) circle (3pt);
	\fill[black] (1.2,-0.5) circle (2pt);
   	\fill[white] (2.2,-0.5) circle (3pt);
	\fill[black] (2.2,-0.5) circle (2pt);
    \draw [dashed] (2.2,0.5) -- (2.7,0);
    \draw [dashed] (2.2,-0.5) -- (2.7,0);
    \draw [decorate,decoration={snake,amplitude=1.5pt,segment length=6pt}] (2.7,0) -- (3.4,0);
\end{tikzpicture}\;.
\label{eq:spectrum_box}
\end{equation}
where the four black dots in the corners of the box can be joined by using solid lines in a number of ways \cite{Li:2023xtl}. The following diagrams violate helicity conservation and therefore vanish after performing the angular integrals, as one can explicitly check
\begin{equation}
\begin{tikzpicture}[baseline={-2}]
    \draw [decorate,decoration={snake,amplitude=1.5pt,segment length=6pt}] (0,0) -- (0.7,0);
    \draw [dashed] (0.7,0) -- (1.2,0.5);
    \draw [dashed] (0.7,0) -- (1.2,-0.5);
\end{tikzpicture}
\quad\Bigg(\quad
\begin{tikzpicture}[baseline={-2}]
	\fill[black] (0,+0.5) circle (2pt);
	\fill[black] (1,+0.5) circle (2pt);
	\fill[black] (0,-0.5) circle (2pt);
	\fill[black] (1,-0.5) circle (2pt);
    \draw (0,-0.5) -- (0,+0.5);
    \draw (0,+0.5) -- (1,+0.5);
    \draw (1,+0.5) -- (1,-0.5);
\end{tikzpicture}
\quad+\quad
\begin{tikzpicture}[baseline={-2}]
	\fill[black] (0,+0.5) circle (2pt);
	\fill[black] (1,+0.5) circle (2pt);
	\fill[black] (0,-0.5) circle (2pt);
	\fill[black] (1,-0.5) circle (2pt);
    \draw (0,-0.5) -- (0,+0.5);
    \draw (0,+0.5) -- (1,+0.5);
    \draw (0,+0.5) -- (1,-0.5);
\end{tikzpicture}
\quad+\quad
\begin{tikzpicture}[baseline={-2}]
	\fill[black] (0,+0.5) circle (2pt);
	\fill[black] (1,+0.5) circle (2pt);
	\fill[black] (0,-0.5) circle (2pt);
	\fill[black] (1,-0.5) circle (2pt);
    \draw (0,-0.5) -- (0.25,0);
    \draw (0,0.5) -- (0.25,0);
    \draw (0.25,0) -- (0.75,0);
    \draw (0.75,0) -- (1,0.5);
    \draw (0.75,0) -- (1,-0.5);
	\fill[black] (0.25,0) circle (2.5pt);
	\fill[white] (0.25,0) circle (1.5pt);
	\fill[black] (0.75,0) circle (2.5pt);
	\fill[white] (0.75,0) circle (1.5pt);
\end{tikzpicture}
\quad+\quad
\begin{tikzpicture}[baseline={-2}]
	\fill[black] (0,+0.5) circle (2pt);
	\fill[black] (1,+0.5) circle (2pt);
	\fill[black] (0,-0.5) circle (2pt);
	\fill[black] (1,-0.5) circle (2pt);
    \draw (0,-0.5) -- (0,+0.5);
    \draw (0,+0.5) -- (1,-0.5);
    \draw (0.5,0) -- (1,0.5);
	\fill[black] (0.5,0) circle (2.5pt);
	\fill[white] (0.5,0) circle (1.5pt);
\end{tikzpicture}
\quad+\quad
\begin{tikzpicture}[baseline={-2}]
	\fill[black] (0,+0.5) circle (2pt);
	\fill[black] (1,+0.5) circle (2pt);
	\fill[black] (0,-0.5) circle (2pt);
	\fill[black] (1,-0.5) circle (2pt);
    \draw (0,+0.5) -- (1,-0.5);
    \draw (0,-0.5) -- (1,+0.5);
	\fill[black] (0.5,0) circle (2.5pt);
	\fill[white] (0.5,0) circle (1.5pt);
\end{tikzpicture}
\quad\Bigg)\quad
\begin{tikzpicture}[baseline={-2}]
    \draw [dashed] (2.2,0.5) -- (2.7,0);
    \draw [dashed] (2.2,-0.5) -- (2.7,0);
    \draw [decorate,decoration={snake,amplitude=1.5pt,segment length=6pt}] (2.7,0) -- (3.4,0);
\end{tikzpicture}
\;.
\end{equation}
Concretely, since ${\bm p}\cdot{\bm e}^s({\bm k})\cdot{\bm p}$ is proportional to $\cos 2\phi_p$ if $s=+$ and to $\sin 2\phi_p$ if $s=\times$, with $\phi_p$ denoting the azimuthal angle of ${\bm p}$ if ${\bm k}$ is taken to be aligned with the $\hat{\bm z}$ axis, any time the two loops in the diagram do not share a propagator the integrand will be independent of $\phi_p$ and the result will vanish. Similarly, diagrams with tadpoles do not contribute to the propagator \cite{Ballesteros:2024zdp}.

There are a total of nine nonvanishing diagrams contributing to the tensor power spectrum. The first five diagrams involve loop corrections to one of the propagators\footnote{The quadratic counterterm also contributes, see the discussion in Section \ref{sec:one_loop_spectrum}.}
\begin{equation}
\begin{tikzpicture}[baseline={-2}]
    \draw [decorate,decoration={snake,amplitude=1.5pt,segment length=6pt}] (0,0) -- (0.7,0);
    \draw [dashed] (0.7,0) -- (1.2,0.5);
    \draw [dashed] (0.7,0) -- (1.2,-0.5);
\end{tikzpicture}
\quad\Bigg(\quad
\begin{tikzpicture}[baseline={-2}]
	\fill[black] (0,+0.5) circle (2pt);
    \draw (0,0.5) to[out=90,in=90] (1,0.5);
    \draw (0,0.5) to[out=-90,in=-90] (1,0.5);
	\fill[black] (1,+0.5) circle (2pt);
	\fill[black] (0,-0.5) circle (2pt);
    \draw (0,-0.5) -- (1,-0.5);
	\fill[black] (1,-0.5) circle (2pt);
\end{tikzpicture}
\quad+\quad
\begin{tikzpicture}[baseline={-2}]
	\fill[black] (0,+0.5) circle (2pt);
    \draw (0,0.25) circle (7pt);
    \draw (0,0.5) -- (1,0.5);
	\fill[black] (1,+0.5) circle (2pt);
	\fill[black] (0,-0.5) circle (2pt);
    \draw (0,-0.5) -- (1,-0.5);
	\fill[black] (1,-0.5) circle (2pt);
\end{tikzpicture}
\quad+\quad
\begin{tikzpicture}[baseline={-2}]
	\fill[black] (0,+0.5) circle (2pt);
    \draw (0.5,0.25) circle (7pt);
    \draw (0,0.5) -- (1,0.5);
	\fill[black] (0.5,0.5) circle (2.5pt);
	\fill[white] (0.5,0.5) circle (1.5pt);
	\fill[black] (1,+0.5) circle (2pt);
	\fill[black] (0,-0.5) circle (2pt);
    \draw (0,-0.5) -- (1,-0.5);
	\fill[black] (1,-0.5) circle (2pt);
\end{tikzpicture}
\quad+\quad
\begin{tikzpicture}[baseline={-2}]
	\fill[black] (0,+0.5) circle (2pt);
    \draw (0.5,0.5) circle (7pt);
    \draw (0,0.5) -- (0.25,0.5);
    \draw (0.75,0.5) -- (1,0.5);
	\fill[black] (0.25,0.5) circle (2.5pt);
	\fill[white] (0.25,0.5) circle (1.5pt);
	\fill[black] (0.75,0.5) circle (2.5pt);
	\fill[white] (0.75,0.5) circle (1.5pt);
	\fill[black] (1,+0.5) circle (2pt);
	\fill[black] (0,-0.5) circle (2pt);
    \draw (0,-0.5) -- (1,-0.5);
	\fill[black] (1,-0.5) circle (2pt);
\end{tikzpicture}
\quad+\quad
\begin{tikzpicture}[baseline={-2}]
	\fill[black] (0,+0.5) circle (2pt);
    \draw (0.25,0.5) circle (7pt);
    \draw (0.5,0.5) -- (1,0.5);
	\fill[black] (0.5,0.5) circle (2.5pt);
	\fill[white] (0.5,0.5) circle (1.5pt);
    \fill[black] (1,+0.5) circle (2pt);
	\fill[black] (0,-0.5) circle (2pt);
    \draw (0,-0.5) -- (1,-0.5);
	\fill[black] (1,-0.5) circle (2pt);
\end{tikzpicture}
\quad\Bigg)\quad
\begin{tikzpicture}[baseline={-2}]
    \draw [dashed] (2.2,0.5) -- (2.7,0);
    \draw [dashed] (2.2,-0.5) -- (2.7,0);
    \draw [decorate,decoration={snake,amplitude=1.5pt,segment length=6pt}] (2.7,0) -- (3.4,0);
\end{tikzpicture}
\;.
\label{eq:loop_prop}
\end{equation}
These diagrams exactly match the calculation in Section \ref{sec:one_loop_spectrum}, so their effect on $\mathcal{P}_h$ can be obtained by replacing one of the $|\varphi_k(t)|^2$ factors by its loop-corrected value in eq.\,(\ref{eq:power_h_tree}) and multiplying the result by $2$ to account for the fact that either the top or bottom line in these diagrams can be corrected.

Aside from the above diagrams which loop-correct one of the propagators, we have another four nonvanishing contributions. There are two diagrams of order $f_{\rm NL}^2$ given by
\begin{align*}
\begin{tikzpicture}[baseline={-2}]
    \draw [decorate,decoration={snake,amplitude=1.5pt,segment length=6pt}] (0,0) -- (0.7,0);
    \draw [dashed] (0.7,0) -- (1.2,0.5);
    \draw [dashed] (0.7,0) -- (1.2,-0.5);
\end{tikzpicture}
\begin{tikzpicture}[baseline={-2}]
	\fill[black] (0,+0.5) circle (2pt);
	\fill[black] (1,+0.5) circle (2pt);
	\fill[black] (0,-0.5) circle (2pt);
	\fill[black] (1,-0.5) circle (2pt);
    \draw (0,+0.5) -- (1,+0.5);
    \draw (0,-0.5) -- (1,-0.5);
    \draw (0,+0.5) -- (0,-0.5);
\end{tikzpicture}
\begin{tikzpicture}[baseline={-2}]
    \draw [dashed] (2.2,0.5) -- (2.7,0);
    \draw [dashed] (2.2,-0.5) -- (2.7,0);
    \draw [decorate,decoration={snake,amplitude=1.5pt,segment length=6pt}] (2.7,0) -- (3.4,0);
\end{tikzpicture}
\quad &\rightarrow \quad
16f_{\rm NL}^2
\int\frac{d^3q}{(2\pi)^3}
\int\frac{d^3p}{(2\pi)^3}
\frac{1}{k^4}
\Big[{\bm p}\cdot {\bm e}^s({\bm k})\cdot{\bm p}\Big]\Big[{\bm q}\cdot {\bm e}^s({\bm k})\cdot{\bm q}\Big]
\\&\hspace{0.9in}
I_k(p,|{\bm k}-{\bm p}|)I_k(q,|{\bm k}+{\bm q}|)
|\varphi_{|{\bm p}+{\bm q}|}(t_e)|^2
|\varphi_q(t_e)|^2
|\varphi_{|{\bm k}+{\bm q}|}(t_e)|^2,
\numberthis
\label{eq:c_diagram}
\\
\begin{tikzpicture}[baseline={-2}]
    \draw [decorate,decoration={snake,amplitude=1.5pt,segment length=6pt}] (0,0) -- (0.7,0);
    \draw [dashed] (0.7,0) -- (1.2,0.5);
    \draw [dashed] (0.7,0) -- (1.2,-0.5);
\end{tikzpicture}
\begin{tikzpicture}[baseline={-2}]
	\fill[black] (0,+0.5) circle (2pt);
	\fill[black] (1,+0.5) circle (2pt);
	\fill[black] (0,-0.5) circle (2pt);
	\fill[black] (1,-0.5) circle (2pt);
    \draw (0,+0.5) -- (1,+0.5);
    \draw (0,-0.5) -- (1,-0.5);
    \draw (1,0.5) -- (0,-0.5);
\end{tikzpicture}
\begin{tikzpicture}[baseline={-2}]
    \draw [dashed] (2.2,0.5) -- (2.7,0);
    \draw [dashed] (2.2,-0.5) -- (2.7,0);
    \draw [decorate,decoration={snake,amplitude=1.5pt,segment length=6pt}] (2.7,0) -- (3.4,0);
\end{tikzpicture}
\quad &\rightarrow \quad
16f_{\rm NL}^2
\int\frac{d^3q}{(2\pi)^3}
\int\frac{d^3p}{(2\pi)^3}
\frac{1}{k^4}
\Big[{\bm p}\cdot {\bm e}^s({\bm k})\cdot{\bm p}\Big]\Big[{\bm q}\cdot {\bm e}^s({\bm k})\cdot{\bm q}\Big]
\\&\hspace{0.9in}
I_k(p,|{\bm k}-{\bm p}|)I_k(q,|{\bm k}+{\bm q}|)
|\varphi_{|{\bm p}+{\bm q}|}(t_e)|^2
|\varphi_q(t_e)|^2
|\varphi_{|{\bm k}-{\bm p}|}(t_e)|^2.
\numberthis
\label{eq:z_diagram}
\end{align*}
There is one mixed diagram with one cubic self interaction and one power of $f_{\rm NL}$, which leads to
\begin{align*}
\begin{tikzpicture}[baseline={-2}]
    \draw [decorate,decoration={snake,amplitude=1.5pt,segment length=6pt}] (0,0) -- (0.7,0);
    \draw [dashed] (0.7,0) -- (1.2,0.5);
    \draw [dashed] (0.7,0) -- (1.2,-0.5);
\end{tikzpicture}
\begin{tikzpicture}[baseline={-2}]
	\fill[black] (0,+0.5) circle (2pt);
	\fill[black] (1,+0.5) circle (2pt);
	\fill[black] (0,-0.5) circle (2pt);
	\fill[black] (1,-0.5) circle (2pt);
    \draw (0,+0.5) -- (1,+0.5);
    \draw (0,+0.5) -- (1,-0.5);
    \draw (0,-0.5) -- (0.5,0);
	\fill[black] (0.5,0) circle (2.5pt);
	\fill[white] (0.5,0) circle (1.5pt);
\end{tikzpicture}
\begin{tikzpicture}[baseline={-2}]
    \draw [dashed] (2.2,0.5) -- (2.7,0);
    \draw [dashed] (2.2,-0.5) -- (2.7,0);
    \draw [decorate,decoration={snake,amplitude=1.5pt,segment length=6pt}] (2.7,0) -- (3.4,0);
\end{tikzpicture}
\quad &\rightarrow \quad
-16f_{\rm NL}
\int\frac{d^3q}{(2\pi)^3}
\int\frac{d^3p}{(2\pi)^3}
\frac{1}{k^4}
\Big[{\bm p}\cdot {\bm e}^s({\bm k})\cdot{\bm p}\Big]\Big[{\bm q}\cdot {\bm e}^s({\bm k})\cdot{\bm q}\Big]
\\&\hspace{0.9in}
I_k(p,|{\bm k}-{\bm p}|)
I_k(q,|{\bm k}+{\bm q}|)
\int_{-\infty}^{t_e} a(t')^3v_3(t') dt'
|\varphi_{|{\bm k}+{\bm q}|}(t_e)|^2
\\&\hspace{0.9in}
2{\rm Im}\Big[
\varphi_q(t_e)\varphi^\star_q(t')
\varphi_{|{\bm p}+{\bm q}|}(t_e)\varphi^\star_{|{\bm p}+{\bm q}|}(t')
\varphi_p(t_e)\varphi^\star_p(t')
\Big].\numberthis
\label{eq:y_diagram}
\end{align*}
Finally, there is a diagram with two cubic self-interactions,
\begin{align*}
\begin{tikzpicture}[baseline={-2}]
    \draw [decorate,decoration={snake,amplitude=1.5pt,segment length=6pt}] (0,0) -- (0.7,0);
    \draw [dashed] (0.7,0) -- (1.2,0.5);
    \draw [dashed] (0.7,0) -- (1.2,-0.5);
\end{tikzpicture}
\begin{tikzpicture}[baseline={-2}]
	\fill[black] (0,+0.5) circle (2pt);
	\fill[black] (1,+0.5) circle (2pt);
	\fill[black] (0,-0.5) circle (2pt);
	\fill[black] (1,-0.5) circle (2pt);
    \draw (0,0.5) -- (0.5,0.25);
    \draw (0.5,0.25) -- (1,0.5);
    \draw (0,-0.5) -- (0.5,-0.25);
    \draw (0.5,-0.25) -- (1,-0.5);
    \draw (0.5,0.25) -- (0.5,-0.25);
	\fill[black] (0.5,0.25) circle (2.5pt);
	\fill[white] (0.5,0.25) circle (1.5pt);
	\fill[black] (0.5,-0.25) circle (2.5pt);
	\fill[white] (0.5,-0.25) circle (1.5pt);
\end{tikzpicture}
\begin{tikzpicture}[baseline={-2}]
    \draw [dashed] (2.2,0.5) -- (2.7,0);
    \draw [dashed] (2.2,-0.5) -- (2.7,0);
    \draw [decorate,decoration={snake,amplitude=1.5pt,segment length=6pt}] (2.7,0) -- (3.4,0);
\end{tikzpicture}
\quad &\rightarrow \quad
4\int\frac{d^3q}{(2\pi)^3}
\int\frac{d^3p}{(2\pi)^3}
\frac{1}{k^4}
\Big[{\bm p}\cdot {\bm e}^s({\bm k})\cdot{\bm p}\Big]\Big[{\bm q}\cdot {\bm e}^s({\bm k})\cdot{\bm q}\Big]
\\&\hspace{0.9in}
I_k(p,|{\bm k}-{\bm p}|)I_k(q,|{\bm k}+{\bm q}|)
\int_{-\infty}^{t_e} a(t')^3v_3(t') dt'
\int_{t'}^{t_e} a(t'')^3v_3(t'') dt''
\\&
\hspace{0.9in}
2{\rm Im}\Big[
\varphi_{|{\bm k}+{\bm q}|}(t_e)
\varphi^\star_{|{\bm k}+{\bm q}|}(t'')
\varphi_{|{\bm k}-{\bm p}|}(t_e)
\varphi^\star_{|{\bm k}-{\bm p}|}(t'')
\Big]\phantom{\int_{-\infty}^\infty}
\\&
\hspace{0.9in}
2{\rm Im}\Big[
\varphi_{q}(t_e)
\varphi^\star_{q}(t')
\varphi_{p}(t_e)
\varphi^\star_{p}(t')
\varphi_{|{\bm p}+{\bm q}|}(t'')
\varphi^\star_{|{\bm p}+{\bm q}|}(t')
\Big].
\numberthis
\label{eq:x_diagram}
\end{align*}
In these expressions we have accounted for the correct symmetry factors and, to be completely clear, the diagrams must be substituted into the box of eq.\,(\ref{eq:spectrum_box}) and the integrals are to be understood as corrections to the right-hand side of eq.\,(\ref{eq:power_h_tree}).

To compute the energy density $\Omega_{\rm GW}$ it is also necessary to take the time average of the $I_k$ terms in the above expressions, which works out to be
\begin{equation}
\frac{k^2}{\mathcal{H}^2}
\overline{
I_k(p,|{\bm k}-{\bm p}|)
I_k(q,|{\bm k}+{\bm q}|)
}
=
\frac{1}{2}\Big[J_s(p,|{\bm k}-{\bm p}|)
J_s(q,|{\bm k}+{\bm q}|)
+
J_c(p,|{\bm k}-{\bm p}|)
J_c(q,|{\bm k}+{\bm q}|)
\Big].
\end{equation}

Before moving on, let us comment on the role of some of the diagrams we have left out of the calculation. Since we assume that the first-order tensor modes are negligible, diagrams with $h$ running in the loops can be safely ignored. In addition to (\ref{eq:spectrum_box}), the one-loop tensor power spectrum also receives a contribution from the diagram
\begin{equation}
\begin{tikzpicture}[baseline={-2}]
    \draw [decorate,decoration={snake,amplitude=1.5pt,segment length=6pt}] (0,0) -- (2.5,0);
    \draw [dashed] (1.25,0.55) circle (15pt);
\end{tikzpicture}\;.
\end{equation}
This diagram was first considered in \cite{Chen:2022dah}, and yields a modulation to the first-order tensor power spectrum. As per our assumption that the leading order tensor modes are negligible, we do not include it. Finally, one could also consider two-loop diagrams with higher-order interaction terms between scalar and tensor modes, such as
\begin{equation}
\begin{tikzpicture}[baseline={-2}]
    \draw [decorate,decoration={snake,amplitude=1.5pt,segment length=6pt}] (0,0) -- (0.7,0);
    \draw [decorate,decoration={snake,amplitude=1.5pt,segment length=6pt}] (1.8,0) -- (2.50,0);    \draw [dashed] (1.25,0) circle (15pt);
    \draw [dashed] (0.7,0) -- (1.8,0);
\end{tikzpicture}\;.
\end{equation}
The reason we have not included these is that they correspond to corrections to the propagation of the tensor modes after the end of inflation (leading, for instance, to additional powers of the scalar transfer functions), whereas we are interested in the correction to the initial conditions. Diagrams of this kind have been considered, for instance, in \cite{Zhou:2021vcw,Zhou:2024ncc}.

\section{Anisotropies}
\label{sec:anis}

In this Section we calculate the anisotropies in the induced gravitational wave background including the intrinsic non-Gaussianity of the inflaton field. We first review the line-of-sight formalism and the solution to the Boltzmann equation, and then proceed to derive an expression for the angular power spectrum with certain approximations. Finally, we compute the angular coefficients by using a diagrammatic approach.

\subsection{The Boltzmann equation}

Let us determine how propagation in an inhomogeneous universe affects the phase space distribution $f(x^\mu,p^\mu)$ of a collection of gravitons, following the treatment of \cite{Contaldi:2016koz,Li:2023xtl,Bartolo:2019oiq,Bartolo:2019yeu,LISACosmologyWorkingGroup:2022kbp}. We can write this quantity in the form $f(\tau,{\bm x},q,\bm n)$, as a function of position $x^\mu$ and momentum $p^\mu=dx^\mu/d\lambda$ for some affine parameter $\lambda$. The direction of propagation is $\bm n={\bm p}/p$ and $q=ap$ denotes the comoving momentum. The distribution obeys the Boltzmann equation
\begin{equation}
\frac{df}{d\lambda}=C[f]+I[f],
\end{equation}
where the $I$ term accounts for graviton emission (from both cosmological and astrophysical sources) and $C$ for their collisions (as discussed in \cite{Contaldi:2016koz}, the line-of-sight formalism presented here is valid in the weak field regime, where this assumption holds, even if the background curvature is large). Since we are interested in the primordial stochastic background, we may set $I=0$ and treat graviton emission as an initial condition. Similarly, graviton self-interactions are Planck-suppressed and can therefore be neglected. We are then interested in solutions to the free Boltzmann equation $df/d\lambda=0$,
\begin{align*}
0=\frac{d\lambda}{d\tau}\frac{df}{d\lambda}
&=
\frac{d\lambda}{d\tau}\bigg(
\frac{\partial f}{\partial x^\mu}p^\mu
+\frac{\partial f}{\partial q}\frac{dq}{d\lambda}
+\frac{\partial f}{\partial n^i}\frac{dn^i}{d\lambda}
\bigg)
\\&=
\frac{\partial f}{\partial \tau}
+\frac{\partial f}{\partial x^i}\frac{dx^i}{d\tau}
+\frac{\partial f}{\partial q}\frac{dq}{d\tau}
+\frac{\partial f}{\partial n^i}\frac{dn^i}{d\tau}.
\numberthis
\label{eq:free_boltzmann}
\end{align*}

We can split the distribution into a background piece which depends only on the comoving momentum $q$ plus a perturbation,\footnote{A thermal bath of bosons, for instance, obeys $f(E,T)=(e^{E/T}-1)^{-1}$ at the background level, which depends only on $q$ because the energy is $E=p=q/a$ and the temperature scales roughly as $T\propto a^{-1}$.}
\begin{align*}
f(\tau,{\bm x},q,\bm n)
&=
\hat{f}(q)+\delta f(\tau,{\bm x},q,\bm n)
\\&\equiv
\hat{f}(q)-q\frac{\partial\hat{f}}{\partial q}\Gamma(\tau,{\bm x},q,\bm n).\numberthis
\label{eq:f_pert}
\end{align*}
We work in the Newtonian gauge assuming no anisotropic stress and ignore vector and tensor perturbations,
\begin{equation}
ds^2=-a^2e^{2\psi}d\tau^2+a^2e^{-2\psi}d{\bm x}^2.
\end{equation}
The geodesic equation yields the following identity \cite{Baumann_2022}
\begin{equation}
\frac{1}{q}\frac{dq}{d\tau}=\frac{\partial\psi}{\partial\tau}-\frac{\partial\psi}{\partial x^i}\frac{dx^i}{d\tau}.
\end{equation}
Since the last term in eq.\,(\ref{eq:free_boltzmann}) is $\mathcal{O}(\psi^2)$, this implies that
\begin{equation}
\frac{\partial \Gamma}{\partial \tau}
+n^i\frac{\partial \Gamma}{\partial x^i}
=
\frac{\partial\psi}{\partial\tau}-n^i\frac{\partial\psi}{\partial x^i}.
\end{equation}
In Fourier space, this equation becomes
\begin{equation}
\frac{\partial \Gamma_{\bm k}}{\partial \tau}
+i({\bm k}\cdot{\bm n})\Gamma_{\bm k}
=
\frac{\partial\psi_{\bm k}}{\partial\tau}-i({\bm k}\cdot{\bm n})\psi_{\bm k},
\end{equation}
where $\Gamma_{\bm k}$ is a function of $(\tau, q, {\bm n})$. This is a sourced linear differential equation with solution
\begin{equation}
\Gamma_{\bm k}(\tau)
=
e^{i{\bm k}\cdot{\bm n}(\tau_\star-\tau)}\Big[\Gamma_{\bm k}(\tau_\star)+\psi_{\bm k}(\tau_\star)\Big]
-\psi_{\bm k}(\tau)
+2\int_{\tau_\star}^\tau d\tau'
\frac{\partial\psi_{\bm k}}{\partial\tau'}e^{i{\bm k}\cdot{\bm n}(\tau'-\tau)}.
\label{eq:gamma_sol}
\end{equation}

The inhomogeneous GW energy density can be obtained by integrating the distribution $f$ over comoving momenta
\begin{equation}
\rho_{\rm GW}({\bm x})= \frac{g}{a^4}\int \frac{d^3q}{(2\pi)^3} qf({\bm x},{\bm q}),
\label{eq:stat_energy}
\end{equation}
where the number of internal degrees of freedom is $g=2$ for gravitons. By comparing this to eq.\,(\ref{eq:gw_energy_fourier}) we can read off the relation
\begin{equation}
f({\bm x},{\bm q})
=
\frac{a^2M_p^2}{32}
\int\frac{d^3p}{(2\pi)^3}e^{i{\bm p}\cdot{\bm x}}
\frac{(q^2-{\bm p}\cdot{\bm q})}{q}
\,{\rm Tr}\Big[
{\bm e}^s({\bm p}-{\bm q})
\cdot
{\bm e}^t({\bm q})
\Big]
\overline{
h_{{\bm p}-{\bm q}}^s
h_{\bm q}^t
}.
\label{eq:f_final}
\end{equation}
It is then straightforward to check that $\hat{f}(q)=\langle f({\bm x},{\bm q})\rangle$ is given by
\begin{equation}
\hat{f}(q)
=
\pi^2
\rho_c
\bigg(\frac{a}{q}\bigg)^4
\Omega_{\rm GW}(q).
\label{eq:f_hat}
\end{equation}

The quantity of interest for anisotropies is the density contrast,
\begin{equation}
\delta_{\rm GW}({\bm x},{\bm q})
=
\frac{f({\bm x},{\bm q})-\hat{f}(q)}{\hat{f}(q)}
=
\bigg[4-\frac{\partial\log\Omega_{\rm GW}(q)}{\partial\log q}\bigg]\Gamma({\bm x},{\bm q}),
\label{eq:delta_gamma}
\end{equation}
which satisfies $\langle\delta_{\rm GW}({\bm x},{\bm k})\rangle=0$.

Following the treatment of \cite{Bartolo:2019zvb,Li:2023xtl}, we neglect both the monopole term $-\psi_{\bm k}(\tau)$ in eq.\,(\ref{eq:gamma_sol}), and the last term, which represents the integrated Sachs-Wolfe effect and has been shown to be subdominant \cite{Bartolo:2019zvb}. Renaming $\tau\leftrightarrow \tau_\star$, we have the relation
\begin{equation}
e^{i{\bm k}\cdot{\bm n}\tau_\star}\Gamma_{\bm k}(\tau_\star)
=
e^{i{\bm k}\cdot{\bm n}\tau}\Big[\Gamma_{\bm k}(\tau)+\psi_{\bm k}(\tau)\Big]
\end{equation}
for an arbitrary time $\tau_\star$, which we take as today. Combining this with eq.\,(\ref{eq:delta_gamma}), we find the density contrast today\footnote{If the tilt of $\Omega_{\rm GW}$ is exactly $4$, then this term vanishes. This is a consequence of the fact that, in this case, $\hat{f}$ in eq.\,(\ref{eq:f_hat}) is independent of $q$ and thus cannot be perturbed. Note, however, that corrections to (\ref{eq:f_final}) of higher-order in perturbations could alter this scaling, and thus terms of order $\sim\Gamma^2$ do not necessarily vanish.}
\begin{align*}
\delta_{\rm GW}(\tau_\star,{\bm x}_\star,{\bm q})
&=
\bigg[4-\frac{\partial\log\Omega_{\rm GW}(\tau,q)}{\partial\log q}\bigg]
\int\frac{d^3k}{(2\pi)^3}
\Big[\Gamma_{\bm k}(\tau,{\bm q})+\psi_{\bm k}(\tau)\Big]
e^{i{\bm k}\cdot[{\bm x}_\star+{\bm n}(\tau-\tau_\star)]}
\\&=
\delta_{\rm GW}(\tau,{\bm x},{\bm q})
+\bigg[4-\frac{\partial\log\Omega_{\rm GW}(\tau,q)}{\partial\log q}\bigg]
\psi(\tau,{\bm x}),\numberthis
\label{eq:present_contrast}
\end{align*}
where ${\bm x}={\bm x}_\star+{\bm n}(\tau-\tau_\star)$.

\subsection{Angular power spectrum}

Anisotropies are typically expressed in terms of the angular power spectrum, the definition and properties of which we now proceed to review. Most of the discussion follows exactly the same logic as for the CMB anisotropies, so we closely follow the discussion in \cite{Baumann_2022}. Let us decompose the density contrast in spherical harmonics,
\begin{equation}
\delta_{\rm GW}(\tau_\star,{\bm x}_\star,{\bm q})
=
\sum_{\ell m}\delta_{\ell m}(\tau_\star,{\bm x}_\star,q)Y_{\ell m}({\bm n}).
\end{equation}
The multipole moments $\delta_{\ell m}(\tau_\star,{\bm x}_\star,q)$ have the following two-point function
\begin{equation}
\langle
\delta_{\ell m}(\tau_\star,{\bm x}_\star,q)
\delta^{\star}_{\ell'm'}(\tau_\star,{\bm x}_\star,q)
\rangle
=
\delta_{\ell\ell'}\delta_{mm'}C_\ell(\tau_\star,{\bm x}_\star,q).
\end{equation}
By using this relation we can find the two-point correlator of $\delta_{\rm GW}$
\begin{align*}
\Delta(\tau_\star,{\bm x}_\star,q,\theta)
&\equiv
\langle
\delta_{\rm GW}(\tau_\star,{\bm x}_\star,q,{\bm n})
\delta_{\rm GW}(\tau_\star,{\bm x}_\star,q,{\bm n}')
\rangle
\phantom{\sum_\ell}
\\&=
\sum_{\ell m}\sum_{\ell' m'}
\langle
\delta_{\ell m}(\tau_\star,{\bm x}_\star,q)\delta^{\star}_{\ell' m'}(\tau_\star,{\bm x}_\star,q)
\rangle
Y_{\ell m}({\bm n})Y^\star_{\ell' m'}({\bm n}')
\\&=
\sum_{\ell}
C_\ell(\tau_\star,{\bm x}_\star,q)
\sum_{m}
Y_{\ell m}({\bm n})Y^\star_{\ell m}({\bm n}')
\\&=
\sum_{\ell}
\frac{2\ell+1}{4\pi}
C_\ell(\tau_\star,{\bm x}_\star,q)
P_\ell(\cos\theta)
\numberthis
\label{eq:angular_pr}
\end{align*}
where $\cos\theta={\bm n}\cdot{\bm n}'$ and we have taken the complex conjugate of $\delta_{\rm GW}(\tau_\star,{\bm x}_\star,q,{\bm n}')$, since it is a real variable.

We can write $\delta_{\rm GW}$ as a Fourier transform
\begin{equation}
\delta_{\rm GW}(\tau_\star,{\bm x}_\star,q,{\bm n})=\int \frac{d^3k}{(2\pi)^3} e^{iD{\bm k}\cdot{\bm n}}\delta_{\rm GW}(\tau_\star,{\bm x}_\star,q,{\bm k}),
\end{equation}
where we have introduced a quantity $D$ with dimensions of distance to make the units consistent. We can expand the plane waves as
\begin{equation}
e^{iD{\bm k}\cdot{\bm n}}
=
\sum_\ell i^\ell(2\ell+1)j_\ell(kD)P_\ell(\hat{\bm k}\cdot{\bm n}),
\end{equation}
where $\hat{\bm k}={\bm k}/k$ and $j_\ell$ is a spherical Bessel function. If we define
\begin{equation}
\langle
\delta_{\rm GW}(\tau_\star,{\bm x}_\star,q,{\bm k})
\delta_{\rm GW}(\tau_\star,{\bm x}_\star,q,{\bm p})
\rangle
=
\frac{2\pi^2}{k^3}\mathcal{P}_{\delta_{\rm GW}}(\tau_\star,{\bm x}_\star,q,k)(2\pi)^3\delta^3({\bm p}+{\bm k}),
\label{eq:delta_twopoint}
\end{equation}
the two-point function becomes
\begin{align*}
\Delta(\tau_\star,{\bm x}_\star,q,\theta)
&=
\int \frac{d^3k}{(2\pi)^3} \sum_\ell i^\ell(2\ell+1)
\int \frac{d^3p}{(2\pi)^3} \sum_{\ell'} i^{\ell'}(2\ell'+1)
\\&
\hspace{1in}
j_\ell(kD)j_{\ell'}(pD)
P_\ell(\hat{\bm k}\cdot{\bm n})
P_{\ell'}(\hat{\bm p}\cdot{\bm n}')
\langle
\delta_{\rm GW}(\tau_\star,{\bm x}_\star,{\bm k})
\delta_{\rm GW}(\tau_\star,{\bm x}_\star,{\bm p})
\rangle
\\&=
\frac{1}{4\pi}
\sum_{\ell\ell'} i^{\ell+\ell'}(2\ell+1)(2\ell'+1)
\\&
\hspace{1in}
\int\frac{dk}{k}
\mathcal{P}_{\delta_{\rm GW}}(\tau_\star,{\bm x}_\star,q,k)
j_\ell(kD)
j_{\ell'}(kD)
\int d\Omega_k
P_\ell(\hat{\bm k}\cdot{\bm n})
P_{\ell'}(-\hat{\bm k}\cdot{\bm n}')
\\&=
\sum_{\ell}\frac{2\ell+1}{4\pi}
\bigg[
4\pi\int\frac{dk}{k}
\mathcal{P}_{\delta_{\rm GW}}(\tau_\star,{\bm x}_\star,q,k)
j^2_\ell(kD)
\bigg]
P_\ell(\cos\theta).
\numberthis
\end{align*}
Comparing this to eq.\,(\ref{eq:angular_pr}), we obtain the following expression for the angular coefficients
\begin{equation}
C_\ell(\tau_\star,{\bm x}_\star,q)
=
4\pi\int\frac{dk}{k}
\mathcal{P}_{\delta_{\rm GW}}(\tau_\star,{\bm x}_\star,q,k)
j^2_\ell(kD).
\label{eq:spherical_bessel}
\end{equation}

\subsection{Diagrammatic calculation}

To find the angular coefficients from eq.\,(\ref{eq:spherical_bessel}) we need to compute the two-point function of $\delta_{\rm GW}$ in (\ref{eq:delta_twopoint}). We see from eqs.\,(\ref{eq:f_final}) and (\ref{eq:delta_gamma}) that each factor of $\delta_{\rm GW}$ contains four copies of $\mathcal{R}$, so this involves computing an eight-point function. We now introduce the Feynman rules corresponding to $\delta_{\rm GW}$,
\begin{equation}
\delta_{\rm GW}
\quad\sim\quad
\begin{tikzpicture}[baseline={-2}]
    \draw [decorate,decoration={snake,amplitude=1pt,segment length=6pt}] (0,0.035) -- (1.5,0.035);
    \draw [decorate,decoration={snake,amplitude=1pt,segment length=6pt}] (0,-0.035) -- (1.5,-0.035);
\end{tikzpicture}\;,
\hspace{0.35in}
\int\frac{d^3p}{(2\pi)^3}e^{i{\bm p}\cdot{\bm x}}
(q^2-{\bm p}\cdot{\bm q})
\,{\rm Tr}\Big[
{\bm e}^s({\bm p}-{\bm q})
\cdot
{\bm e}^t({\bm q})
\Big]
\quad\sim\quad
\begin{tikzpicture}[baseline={-2}]
    \draw [decorate,decoration={snake,amplitude=1pt,segment length=6pt}] (0,0.035) -- (0.7,0.035);
    \draw [decorate,decoration={snake,amplitude=1pt,segment length=6pt}] (0,-0.035) -- (0.7,-0.035);
    \draw [decorate,decoration={snake,amplitude=1.5pt,segment length=6pt}] (0.7,0) -- (1.2,0.5);
    \draw [decorate,decoration={snake,amplitude=1.5pt,segment length=6pt}] (0.7,0) -- (1.2,-0.5);
\end{tikzpicture}\;.
\end{equation}

If we consider only the Gaussian contributions to the two-point function of $\delta_{\rm GW}$, we find the following connected diagrams
\begin{equation}
\begin{tikzpicture}[baseline={-2}]
    \draw [decorate,decoration={snake,amplitude=1pt,segment length=6pt}] (-0.5,0.035) -- (0.3,0.035);
    \draw [decorate,decoration={snake,amplitude=1pt,segment length=6pt}] (-0.5,-0.035) -- (0.3,-0.035);
    \draw [decorate,decoration={snake,amplitude=1.5pt,segment length=6pt}] (0.3,0) to[out=90,in=180] (1.1,0.7);
    \draw [decorate,decoration={snake,amplitude=1.5pt,segment length=6pt}] (0.3,0) to[out=-90,in=180] (1.1,-0.7);
    \draw [dashed] (1.1,+0.7) -- (2.5,+0.7);
    \draw [dashed] (1.1,+0.7) to[out=-90,in=-90] (2.5,+0.7);
    \draw [dashed] (1.1,-0.7) -- (2.5,-0.7);
    \draw [dashed] (1.1,-0.7) to[out=90,in=90] (2.5,-0.7);
    \draw [decorate,decoration={snake,amplitude=1.5pt,segment length=6pt}] (2.5,0.7) to[out=0,in=90] (3.3,0);
    \draw [decorate,decoration={snake,amplitude=1.5pt,segment length=6pt}] (2.5,-0.7) to[out=0,in=-90] (3.3,0);
    \draw [decorate,decoration={snake,amplitude=1pt,segment length=6pt}] (4.1,0.035) -- (3.3,0.035);
    \draw [decorate,decoration={snake,amplitude=1pt,segment length=6pt}] (4.1,-0.035) -- (3.3,-0.035);
\end{tikzpicture}
\; + \;
\begin{tikzpicture}[baseline={-2}]
    \draw [decorate,decoration={snake,amplitude=1pt,segment length=6pt}] (-0.5,0.035) -- (0.3,0.035);
    \draw [decorate,decoration={snake,amplitude=1pt,segment length=6pt}] (-0.5,-0.035) -- (0.3,-0.035);
    \draw [decorate,decoration={snake,amplitude=1.5pt,segment length=6pt}] (0.3,0) to[out=90,in=180] (1.1,0.7);
    \draw [decorate,decoration={snake,amplitude=1.5pt,segment length=6pt}] (0.3,0) to[out=-90,in=180] (1.1,-0.7);
    \draw [dashed] (1.1,+0.7) -- (2.5,+0.7);
    \draw [dashed] (1.1,+0.7) -- (2.5,-0.7);
    \draw [dashed] (1.1,-0.7) -- (2.5,-0.7);
    \draw [dashed] (1.1,-0.7) -- (2.5,+0.7);
    \draw [decorate,decoration={snake,amplitude=1.5pt,segment length=6pt}] (2.5,0.7) to[out=0,in=90] (3.3,0);
    \draw [decorate,decoration={snake,amplitude=1.5pt,segment length=6pt}] (2.5,-0.7) to[out=0,in=-90] (3.3,0);
    \draw [decorate,decoration={snake,amplitude=1pt,segment length=6pt}] (4.1,0.035) -- (3.3,0.035);
    \draw [decorate,decoration={snake,amplitude=1pt,segment length=6pt}] (4.1,-0.035) -- (3.3,-0.035);
\end{tikzpicture}
\; + \;
\begin{tikzpicture}[baseline={-2}]
    \draw [decorate,decoration={snake,amplitude=1pt,segment length=6pt}] (-0.5,0.035) -- (0.3,0.035);
    \draw [decorate,decoration={snake,amplitude=1pt,segment length=6pt}] (-0.5,-0.035) -- (0.3,-0.035);
    \draw [decorate,decoration={snake,amplitude=1.5pt,segment length=6pt}] (0.3,0) to[out=90,in=180] (1.1,0.7);
    \draw [decorate,decoration={snake,amplitude=1.5pt,segment length=6pt}] (0.3,0) to[out=-90,in=180] (1.1,-0.7);
    \draw [dashed] (1.1,+0.7) -- (2.5,+0.7);
    \draw [dashed] (1.1,+0.7) -- (1.1,-0.7);
    \draw [dashed] (1.1,-0.7) -- (2.5,-0.7);
    \draw [dashed] (2.5,+0.7) -- (2.5,-0.7);
    \draw [decorate,decoration={snake,amplitude=1.5pt,segment length=6pt}] (2.5,0.7) to[out=0,in=90] (3.3,0);
    \draw [decorate,decoration={snake,amplitude=1.5pt,segment length=6pt}] (2.5,-0.7) to[out=0,in=-90] (3.3,0);
    \draw [decorate,decoration={snake,amplitude=1pt,segment length=6pt}] (4.1,0.035) -- (3.3,0.035);
    \draw [decorate,decoration={snake,amplitude=1pt,segment length=6pt}] (4.1,-0.035) -- (3.3,-0.035);
\end{tikzpicture}\;.
\label{eq:nofnl_diagrams}
\end{equation}
As shown in \cite{Bartolo:2019zvb}, however, these diagrams yield highly suppressed contributions to the anisotropy at low $\ell$ due to the fact that GWs are induced locally, and therefore patches in the sky separated by large distances are essentially uncorrelated.

Non-Gaussianities present us with a way out of this problem since, in the presence of a non-Gaussian interaction, two short-wavelength modes may conspire to create a long-wavelength one, thereby introducing a correlation between distant patches \cite{Bartolo:2019zvb}. If we split the linear curvature perturbation $\varphi$ into long- and short-wavelength modes
\begin{equation}
\varphi=\varphi_{\rm S}+\varphi_{\rm L},
\end{equation}
and assume, for illustrative purposes, a non-Gaussianity of the local type $\mathcal{R}=\varphi+f_{\rm NL}\varphi^2$, then we have
\begin{equation}
\mathcal{R}^4
=
\varphi_{\rm S}^4+\mathcal{O}(\varphi_{\rm S}^5)
+\varphi_{\rm L}\Big[4\varphi_{\rm S}^3+20f_{\rm NL}\varphi_{\rm S}^4+\mathcal{O}(\varphi_{\rm S}^5)\Big]
+\mathcal{O}(\varphi_{\rm L}^2).
\end{equation}
Upon computing the eight-point function $\langle\mathcal{R}^8\rangle$, since long- and short-wavelength modes are Gaussian and uncorrelated, we would find contributions $\langle\varphi_{\rm S}^8\rangle$ and $\langle\varphi_{\rm L}^2\rangle\langle\varphi_{\rm S}^6\rangle$ already included in the three diagrams of eq.\,(\ref{eq:nofnl_diagrams}) which, as we have mentioned, are highly suppressed at low $\ell$, and a contribution $f_{\rm NL}\langle\varphi_{\rm L}^2\rangle\langle\varphi_{\rm S}^8\rangle$ corresponding to the diagram
\begin{equation}
\begin{tikzpicture}[baseline={-2}]
    \draw [decorate,decoration={snake,amplitude=1pt,segment length=6pt}] (-0.5,0.035) -- (0.3,0.035);
    \draw [decorate,decoration={snake,amplitude=1pt,segment length=6pt}] (-0.5,-0.035) -- (0.3,-0.035);
    \draw [decorate,decoration={snake,amplitude=1.5pt,segment length=6pt}] (0.3,0) to[out=90,in=180] (1.1,0.7);
    \draw [decorate,decoration={snake,amplitude=1.5pt,segment length=6pt}] (0.3,0) to[out=-90,in=180] (1.1,-0.7);
    \draw [dashed] (1.1,+0.7) -- (1.45,+0.35);
    \draw [dashed] (1.1,+0.7) -- (0.75,+0.35);
    \draw [dashed] (1.1,-0.7) -- (1.45,-0.35);
    \draw [dashed] (1.1,-0.7) -- (0.75,-0.35);
    \draw (0.75,+0.35) -- (0.75,-0.35);
    \draw (1.45,+0.35) -- (1.45,-0.35);
	\fill[black] (1.45,+0.35) circle (2pt);
	\fill[black] (1.45,-0.35) circle (2pt);
	\fill[black] (0.75,+0.35) circle (2pt);
	\fill[black] (0.75,-0.35) circle (2pt);
    \draw (1.45,0.39) -- (2.15,0.39);
    \draw (1.45,0.31) -- (2.15,0.31);
    \draw (2.15,+0.35) -- (2.15,-0.35);
    \draw (2.85,+0.35) -- (2.85,-0.35);
	\fill[black] (2.15,+0.35) circle (2pt);
	\fill[black] (2.15,-0.35) circle (2pt);
	\fill[black] (2.85,+0.35) circle (2pt);
	\fill[black] (2.85,-0.35) circle (2pt);
    \draw [dashed] (2.5,+0.7) -- (2.15,+0.35);
    \draw [dashed] (2.5,+0.7) -- (2.85,+0.35);
    \draw [dashed] (2.5,-0.7) -- (2.15,-0.35);
    \draw [dashed] (2.5,-0.7) -- (2.85,-0.35);
    \draw [decorate,decoration={snake,amplitude=1.5pt,segment length=6pt}] (2.5,0.7) to[out=0,in=90] (3.3,0);
    \draw [decorate,decoration={snake,amplitude=1.5pt,segment length=6pt}] (2.5,-0.7) to[out=0,in=-90] (3.3,0);
    \draw [decorate,decoration={snake,amplitude=1pt,segment length=6pt}] (4.1,0.035) -- (3.3,0.035);
    \draw [decorate,decoration={snake,amplitude=1pt,segment length=6pt}] (4.1,-0.035) -- (3.3,-0.035);
\end{tikzpicture}
\end{equation}
which is not included in (\ref{eq:nofnl_diagrams}) and is not suppressed at low $\ell$, with the price to pay being the small factor of the long-wavelength power spectrum $\langle\varphi_{\rm L}^2\rangle\sim 10^{-9}$ multiplying the result, represented by the double line
\begin{equation}
\varphi_{\rm L}
\quad\sim\quad
\begin{tikzpicture}
    \draw (0,0.54) -- (1.5,0.54);
    \draw (0,0.46) -- (1.5,0.46);
\end{tikzpicture}\;.
\end{equation}
This line was dubbed a {\it non-Gaussian bridge} in \cite{Bartolo:2019zvb,Li:2023xtl}, since it connects both sides of the previous diagram.

Other non-Gaussian bridges are possible once we consider $g_{\rm NL}$ and higher-order terms as well as self-interactions. If we restrict ourselves to a maximum of four loops (two on each side) as in the above diagram, two more bridges involving a cubic self-interaction are possible. One of them is
\begin{equation}
\begin{tikzpicture}[baseline={-2}]
    \draw [decorate,decoration={snake,amplitude=1pt,segment length=6pt}] (-0.5,0.035) -- (0.3,0.035);
    \draw [decorate,decoration={snake,amplitude=1pt,segment length=6pt}] (-0.5,-0.035) -- (0.3,-0.035);
    \draw [decorate,decoration={snake,amplitude=1.5pt,segment length=6pt}] (0.3,0) to[out=90,in=180] (1.1,0.7);
    \draw [decorate,decoration={snake,amplitude=1.5pt,segment length=6pt}] (0.3,0) to[out=-90,in=180] (1.1,-0.7);
    \draw [dashed] (1.1,+0.7) -- (1.45,+0.35);
    \draw [dashed] (1.1,+0.7) -- (0.75,+0.35);
    \draw [dashed] (1.1,-0.7) -- (1.45,-0.35);
    \draw [dashed] (1.1,-0.7) -- (0.75,-0.35);
    \draw (0.75,+0.35) -- (1.45,-0.35);
    \draw (0.75,-0.35) -- (1.1,0);
	\fill[black] (1.45,+0.35) circle (2pt);
	\fill[black] (1.45,-0.35) circle (2pt);
	\fill[black] (0.75,+0.35) circle (2pt);
	\fill[black] (0.75,-0.35) circle (2pt);
    \draw (1.45,0.39) -- (2.15,0.39);
    \draw (1.45,0.31) -- (2.15,0.31);
	\fill[black] (1.1,0) circle (2.5pt);
	\fill[white] (1.1,0) circle (1.5pt);
\end{tikzpicture}
\end{equation}
which vanishes once we take the outgoing field to be long-wavelength. The second possible bridge involving a self-interaction does not vanish, and adding the two diagrams together we find, schematically,
\begin{equation}
\langle\delta_{\rm GW}^2\rangle
\quad\sim\quad
\Bigg(\;
\begin{tikzpicture}[baseline={-2}]
    \draw [decorate,decoration={snake,amplitude=1pt,segment length=6pt}] (-0.5,0.035) -- (0.3,0.035);
    \draw [decorate,decoration={snake,amplitude=1pt,segment length=6pt}] (-0.5,-0.035) -- (0.3,-0.035);
    \draw [decorate,decoration={snake,amplitude=1.5pt,segment length=6pt}] (0.3,0) to[out=90,in=180] (1.1,0.7);
    \draw [decorate,decoration={snake,amplitude=1.5pt,segment length=6pt}] (0.3,0) to[out=-90,in=180] (1.1,-0.7);
    \draw [dashed] (1.1,+0.7) -- (1.45,+0.35);
    \draw [dashed] (1.1,+0.7) -- (0.75,+0.35);
    \draw [dashed] (1.1,-0.7) -- (1.45,-0.35);
    \draw [dashed] (1.1,-0.7) -- (0.75,-0.35);
    \draw (0.75,+0.35) -- (0.75,-0.35);
    \draw (1.45,+0.35) -- (1.45,-0.35);
	\fill[black] (1.45,+0.35) circle (2pt);
	\fill[black] (1.45,-0.35) circle (2pt);
	\fill[black] (0.75,+0.35) circle (2pt);
	\fill[black] (0.75,-0.35) circle (2pt);
    \draw (1.45,0.39) -- (2.15,0.39);
    \draw (1.45,0.31) -- (2.15,0.31);
\end{tikzpicture}
\; + \;
\begin{tikzpicture}[baseline={-2}]
    \draw [decorate,decoration={snake,amplitude=1pt,segment length=6pt}] (-0.5,0.035) -- (0.3,0.035);
    \draw [decorate,decoration={snake,amplitude=1pt,segment length=6pt}] (-0.5,-0.035) -- (0.3,-0.035);
    \draw [decorate,decoration={snake,amplitude=1.5pt,segment length=6pt}] (0.3,0) to[out=90,in=180] (1.1,0.7);
    \draw [decorate,decoration={snake,amplitude=1.5pt,segment length=6pt}] (0.3,0) to[out=-90,in=180] (1.1,-0.7);
    \draw [dashed] (1.1,+0.7) -- (1.45,+0.35);
    \draw [dashed] (1.1,+0.7) -- (0.75,+0.35);
    \draw [dashed] (1.1,-0.7) -- (1.45,-0.35);
    \draw [dashed] (1.1,-0.7) -- (0.75,-0.35);
    \draw (0.75,+0.35) -- (0.75,-0.35);
    \draw (1.45,+0.35) -- (1.45,-0.35);
	\fill[black] (1.45,+0.35) circle (2pt);
	\fill[black] (1.45,-0.35) circle (2pt);
	\fill[black] (0.75,+0.35) circle (2pt);
	\fill[black] (0.75,-0.35) circle (2pt);
    \draw (1.45,0.04) -- (2.15,0.04);
    \draw (1.45,-0.04) -- (2.15,-0.04);
	\fill[black] (1.45,0) circle (2.5pt);
	\fill[white] (1.45,0) circle (1.5pt);
\end{tikzpicture}
\;\Bigg)^2.
\label{eq:delta_diagrams}
\end{equation}

It is clear from these diagrams that it should be possible to factor a long-wavelength field from the density contrast, $\delta_{\rm GW}\propto \varphi_{\rm L}$. Let us write
\begin{equation}
\delta_{\rm GW}(\tau,{\bm x},q,{\bm k})
=
\frac{\Omega_{\rm NG}(\tau,q)}{\Omega_{\rm GW}(\tau,q)}
\varphi^{\rm L}_{\bm k}(\tau),
\end{equation}
where $\Omega_{\rm NG}$ is a proportionality factor that, roughly speaking, corresponds to the middle section of the above diagrams and could, in principle, depend on ${\bm x}$, but this dependence drops out once we consider the long-wavelength limit ${\bm k}\rightarrow 0$.\footnote{In fact, this is the reason these diagrams are unsuppressed with respect to those in (\ref{eq:nofnl_diagrams}). For the latter diagrams the spatial dependence leads to a suppression $\langle\delta_{\rm GW}({\bm x})\delta_{\rm GW}({\bm y})\rangle\propto (k_\star|{\bm x}-{\bm y}|)^{-3}$, where $k_\star$ is the location of the peak in the scalar power spectrum \cite{Bartolo:2019zvb}.}

Before computing $\Omega_{\rm NG}$, let us assume this expression is valid and compute the angular coefficients. Only the long-wavelength part of $\psi$ in eq.\,(\ref{eq:present_contrast}) contributes to the correlator $\langle\delta_{\rm GW}^2\rangle$. Using eq.\,(\ref{eq:newton_initial}) and assuming the long modes re-enter during the matter-dominated era, the Newtonian potential can be written as
\begin{equation}
\phi_{\bm k}(\tau)
=
\frac{3}{5}\varphi^{\rm L}_{\bm k}.
\end{equation}
The power spectrum of $\delta_{\rm GW}$ is therefore
\begin{equation}
\mathcal{P}_{\delta_{\rm GW}}(\tau,{\bm x},q,k)
=
\bigg\{
\frac{\Omega_{\rm NG}(\tau,{\bm x},q)}{\Omega_{\rm GW}(\tau,q)}
+\frac{3}{5}\bigg[4-\frac{\partial\log\Omega_{\rm GW}(\tau,q)}{\partial\log q}\bigg]
\bigg\}^2
\mathcal{P}^{\rm L}_\mathcal{R}(k).
\end{equation}
Using the fact that $\mathcal{P}^{\rm L}_\mathcal{R}(k)\simeq \mathcal{P}^{\rm L}_\mathcal{R}\sim 10^{-9}$ is approximately scale invariant in eq.\,(\ref{eq:spherical_bessel}), together with the identity
\begin{equation}
\int_0^\infty j_\ell^2(x)\frac{dx}{x}=\frac{1}{2\ell(\ell+1)},
\end{equation}
we find \cite{Bartolo:2019zvb,Li:2023xtl}
\begin{equation}
C_\ell(\tau_\star,x_\star,q)
=
\frac{2\pi\mathcal{P}^{\rm L}_\mathcal{R}}{\ell(\ell+1)}
\bigg\{
\frac{\Omega_{\rm NG}(\tau_\star,{\bm x}_\star,q)}{\Omega_{\rm GW}(\tau_\star,q)}
+\frac{3}{5}\bigg[4-\frac{\partial\log\Omega_{\rm GW}(\tau_\star,q)}{\partial\log q}\bigg]
\bigg\}^2.
\label{eq:anisotropies}
\end{equation}
Note that the combination $\ell(\ell+1)C_\ell(\tau_\star,{\bm x}_\star,q)$ is independent of $\ell$. The first term in this expression accounts for anisotropies in the gravitational wave production, and the second one for the anisotropies accumulated when the tensor modes propagate in an inhomogeneous Universe.

To find $\Omega_{\rm NG}$, we simply need to calculate the correlator $\langle \delta_{\rm GW}({\bm x},{\bm q})\delta_{\rm GW}({\bm y},{\bm q})\rangle$. This can be done by using eq.\,(\ref{eq:h_solution}) to write the inhomogeneous energy density in eq.\,(\ref{eq:gw_energy_fourier}) in terms of $\mathcal{R}$,\footnote{We have omitted the dependence of $I$ on $\tau$ on the right-hand side, since it disappears after averaging and taking the late-time limit.}
\begin{align*}
\Omega_{\rm GW}(\tau,&{\bm x},{\bm q})
=
\frac{1}{48a^2H^2}\frac{q^3}{2\pi^2}
\int\frac{d^3p}{(2\pi)^3}
\int\frac{d^3k}{(2\pi)^3}
\int\frac{d^3\ell}{(2\pi)^3}
\frac{(q^2-{\bm p}\cdot{\bm q})}{q^2|{\bm p}-{\bm q}|^2}
e^{i{\bm p}\cdot {\bm x}}
\mathcal{R}_{\bm k}\mathcal{R}_{{\bm p}-{\bm q}-{\bm k}}
\mathcal{R}_{\bm \ell}\mathcal{R}_{{\bm q}-{\bm \ell}}
\\&
{\rm Tr}\Big[{\bm e}^s({\bm p}-{\bm q})\cdot{\bm e}^t({\bm q})\Big]
\Big[{\bm k}\cdot {\bm e}^s({\bm p}-{\bm q})\cdot{\bm k}\Big]
\Big[{\bm \ell}\cdot {\bm e}^t({\bm q})\cdot{\bm \ell}\Big]
\overline{
I_{|{\bm p}-{\bm q}|}(k,|{\bm p}-{\bm q}-{\bm k}|)
I_q(\ell,|{\bm q}-{\bm \ell}|)
}.\numberthis
\end{align*}
Using the diagrams in (\ref{eq:delta_diagrams}) to contract the fields appropriately, extracting the long-wavelength field $\varphi_{\bm p}^{\rm L}$ and taking ${\bm p}\rightarrow 0$ in the final result (which eliminates the exponential and makes the result space-independent), we obtain
\begin{align*}
\Omega_{\rm NG}(\tau,q)
&=
\frac{1}{48a^2H^2}
\frac{q^3}{2\pi^2}
\int\frac{d^3p}{(2\pi)^3}
\frac{1}{q^2}
\Big[{\bm p}\cdot {\bm e}^s({\bm q})\cdot{\bm p}\Big]^2
\overline{
I_q(p,|{\bm q}-{\bm p}|)^2
}
\\&
\hspace{0.5in}
|\varphi_{|{\bm q}-{\bm p}|}(t_e)|^2
\bigg\{
16 f_{\rm NL}|\varphi_p(t_e)|^2
-
4\int_{-\infty}^{t_e} dt' a(t')^3 v_3(t') 2{\rm Im}
\Big[
\varphi_{p}(t_e)^2\varphi^\star_{p}(t')^2
\Big]
\bigg\},\numberthis
\end{align*}
where we have also accounted for the symmetry factors of each diagram. This equation can be written in a simpler form by noting that the quantity in brackets is precisely the bispectrum of $\mathcal{R}$ in the squeezed limit, so the consistency relation \cite{Maldacena:2002vr,Creminelli:2004yq} can be used. The bispectrum $B_\mathcal{R}$ in the squeezed limit $k\ll p$ is given by
\begin{equation}
\frac{B_{\mathcal{R}}(p,p,k)}{
|\varphi_k(t_e)|^2
|\varphi_p(t_e)|^2
}
=
4f_{\rm NL}
-
\frac{1}{|\varphi_p(t_e)|^2}\int_{-\infty}^{t_e} dt' a(t')^3 v_3(t')
\;2{\rm Im}
\Big[
\varphi_p(t_e)^2\varphi^\star_p(t')^2
\Big]
=
-\frac{d\log \mathcal{P}_\mathcal{R}(p)}{d\log p},
\end{equation}
where the last equality is due to the consistency relation. Thus,
\begin{equation}
\Omega_{\rm NG}(t,q)
=
-\frac{2}{24}\frac{q^2}{\mathcal{H}^2}
\frac{q^3}{2\pi^2}
\int\frac{d^3p}{(2\pi)^3}
\frac{1}{q^4}
\Big[{\bm p}\cdot {\bm e}^s({\bm q})\cdot{\bm p}\Big]^2
\overline{
I_q(p,|{\bm q}-{\bm p}|)^2
}
|\varphi_{|{\bm q}-{\bm p}|}(t_e)|^2
|\varphi_p(t_e)|^2
\frac{d\log \mathcal{P}_\mathcal{R}(p)}{d\log p}.
\label{eq:omega_ng}
\end{equation}
This expression for the induced gravitational wave anisotropies in terms of the scalar spectral index on small scales is shown here for the first time.

\section{Integral results}
\label{sec:numerical}

In this Section we describe an analytical model for the evolution of the inflaton perturbation first presented in \cite{Ballesteros:2024zdp} that captures the dynamics of the smoothness and duration of the transitions in and out of USR and simplifies the calculation of the time integrals in the in-in formalism. We first discuss how to deal with the divergences in the different diagrams and then present our numerical results.

\subsection{Structure of the divergences}

The equation of motion for the field perturbation $\delta\phi_k$ is, in the $\delta\phi$ gauge,
\begin{equation}
\delta\phi''_k
+2aH\delta\phi'_k
+(k^2+a^2V_2)\delta\phi_k
=
0.
\label{eq:delta_phi_eom}
\end{equation}
The second and third derivatives of the potential can be written as
\begin{equation}
a^2V_2=-a^2H^2\bigg(\nu^2-\frac{9}{4}\bigg),
\qquad
a^2V_3=-\frac{aH}{\sqrt{2\epsilon}}(\nu^2)',
\end{equation}
where, in the limit $\epsilon\ll 1$,
\begin{equation}
\nu^2
=
\frac{9}{4}
-\bigg(
3\eta-\eta^2+\frac{\eta'}{aH}
\bigg).
\end{equation}

Eq.\,(\ref{eq:delta_phi_eom}) can be solved analytically if $\nu^2$ is taken as a piecewise-constant function and we impose Bunch-Davies initial conditions \cite{Ballesteros:2024zdp}. We model the field evolution by using a sequence of five phases of constant (positive or negative) $\nu^2$. In the first phase the field is in slow-roll with $\nu=3/2$. During the second phase we have $\nu^2<0$ (and therefore imaginary $\nu$) so that $\eta$ increases to a large, positive value, leading to an ultra-slow-roll phase with constant $\eta\gtrsim 3/2$. During the fourth phase, $\nu^2$ takes a large, positive value so that $\eta$ decreases until the final constant-roll phase in which $\eta<0$.\footnote{In \cite{Ballesteros:2024zdp}, the final stage was taken as a second slow-roll phase instead.} The solution to eq.\,(\ref{eq:delta_phi_eom}) is then
\begin{equation}
\delta\phi_k
=
(-k\tau)^{3/2}\Big[\alpha_kJ_\nu(-k\tau)+\beta_kY_\nu(-k\tau)\Big],
\end{equation}
where $J$ and $Y$ are Bessel functions and the coefficients $\alpha_k$ and $\beta_k$ can be found by imposing continuity of the solutions and their derivatives between transitions. We have taken $H$ as constant for simplicity, and $aH=-1/\tau$. We denote the duration of the USR phase as $\Delta N$, and the duration of the transitions in and out of USR, which we take to be equal, as $\delta N$. The values of $\eta$ during the USR phase and the subsequent CR phase are denoted by $\eta_{\rm USR}$ and $\eta_{\rm CR}$, respectively. In our numerical calculations we take $\epsilon=10^{-3}$ in the far past, and $H^2=8\pi^2\epsilon A_s$ with $A_s=2.2\times 10^{-9}$. The evolution of the slow-roll parameters in shown in Fig.\,\ref{fig:pars_ref}.

This model has two significant advantages. One is that it allows us to control the smoothness of the transitions in and out of USR and the total duration of the phase. Another one is that it allows us to perform the time integrals in all loop calculations immediately. By writing out the piecewise $\nu^2$ in terms of step functions, it is straightforward to show that
\begin{equation}
(\nu^2)'
=
\sum_{i=1}^4
(\nu^2_{i+1}-\nu^2_i)\delta(\tau-\tau_i)
=
\sum_{i=1}^4
\Delta\nu_i^2\delta(\tau-\tau_i),
\end{equation}
so that $v_3=-2\epsilon H(\nu^2)'/a$ can be written as a sum of Dirac delta functions. 

The integrals corresponding to eqs.\,(\ref{eq:fg_loops_1}) and (\ref{eq:fg_loops_2}) are divergent in both the IR and UV. We deal with this by imposing cutoffs, integrating only the region around the peak of the power spectrum. This is similar in spirit to what is done in \cite{Li:2023xtl}, where a lognormal power spectrum that quickly decays in both limits is considered, making the integrals convergent.\footnote{We remark that these divergences are in a different footing to those arising from the in-in formalism, since they are related to the fact that one is attempting to compute statistical properties of the field, such as the variance, in an infinite volume, and persist even if the fields are not quantum in nature. Imposing cutoffs is, roughly speaking, equivalent to stating that any experiment used to measure these properties would have access only to a finite range of momenta, so these divergences would never show up in an actual observable. As a straightforward way of understanding this, one could imagine setting up a lattice simulation of a random Gaussian field and measuring the corresponding variance, which would correspond to integrating the power spectrum between the natural momentum cutoffs given by the lattice size and spacing, and would therefore be completely finite.\label{fn:divergences}
} The integrals in eqs.\,(\ref{eq:cubic_loop_1}), (\ref{eq:cubic_loop_2}) and (\ref{eq:quartic_loop}) require a more delicate treatment, due to the time dependence. The integral (\ref{eq:quartic_loop}) is completely reabsorbed in the counterterm and only yields a finite contribution to the spectrum upon imposing renormalization conditions, so we do not include it in our analysis.

Let us first deal with eq.\,(\ref{eq:cubic_loop_1}). The time integral can be performed immediately by using the analytical model. We obtain
\begin{align*}
\begin{tikzpicture}[baseline={-2}]
    \draw [dashed] (0,0) -- (0.55,0);
    \draw [dashed] (2,0) -- (2.5,0);
    \draw (1.25,0) -- (2,0);
    \draw (0.9,0) circle (10pt);
	\fill[black] (2,0) circle (2pt);
    \fill[black] (0.55,0) circle (2pt);
	\fill[black] (1.25,0) circle (2.5pt);
	\fill[white] (1.25,0) circle (1.5pt);
\end{tikzpicture}
\quad &= \quad
2f_{\rm NL}H
\sum_{i=1}^4
a(\tau_i)^3
2\epsilon(\tau_i)
\Delta\nu_i^2
\\&
\hspace{1in}
\int\frac{d^3p}{(2\pi)^3}
2{\rm Im}\Big[
\varphi_{|{\bm k}-{\bm p}|}(\tau)
\varphi^\star_{|{\bm k}-{\bm p}|}(\tau_i)
\varphi_k(\tau)
\varphi^\star_k(\tau_i)
\varphi_p(\tau)
\varphi^\star_p(\tau_i)
\Big].
\numberthis
\label{eq:anal_cubic_1}
\end{align*}
For eq.\,(\ref{eq:cubic_loop_2}), the situation is more complicated. As shown in \cite{Ballesteros:2024zdp}, to remove the UV divergence in this integral it is necessary to keep the $i\omega$ prescription\footnote{This prescription is used to turn off the interactions in the far past and is implemented by thinking of time as a complex number and choosing the integral contours appropriately.} in $\tau'$ (see the corresponding discussion around eq.\,(\ref{eq:iwpresc}) below) and note that the UV cutoff restricts the domains of the time integrals as
\begin{equation}
\int_{-\infty}^\tau d\tau'
\int_{\tau'}^\tau d\tau''
\quad\rightarrow\quad
\int_{-\infty}^{\tau+1/a(\tau)\Lambda_{\rm UV}} d\tau'
\int_{\tau'+1/a(\tau')\Lambda_{\rm UV}}^{\tau} d\tau''.
\end{equation}
In principle, due to the Dirac deltas coming from the two vertices, the integrals can be split into a sum of sixteen distinct contributions. However, even without including the cutoff in the above integral, six of these points are removed because the $\tau''$ integral does not include points with $\tau''<\tau'$. Including the effect of the cutoff, we find that the four points with $\tau''=\tau'$ do not contribute to the integral either, and only the final six terms remain. This translates into the following sum
\begin{align*}
\begin{tikzpicture}[baseline={-2}]
    \draw [dashed] (0,0) -- (0.5,0);
    \draw (0.5,0) -- (0.9,0);
    \draw [dashed] (2,0) -- (2.5,0);
    \draw (1.6,0) -- (2,0);
    \draw (1.25,0) circle (10pt);
	\fill[black] (0.5,0) circle (2pt);
	\fill[black] (2,0) circle (2pt);
    \fill[black] (0.9,0) circle (2.5pt);
	\fill[white] (0.9,0) circle (1.5pt);
	\fill[black] (1.6,0) circle (2.5pt);
	\fill[white] (1.6,0) circle (1.5pt);
\end{tikzpicture}
\quad &= \quad
H^2\sum_{i=1}^3\sum_{j=i+1}^4
\Big[a(\tau_i)^32\epsilon(\tau_i)\Delta\nu_i^2\Big]
\Big[a(\tau_j)^32\epsilon(\tau_j)\Delta\nu_j^2\Big]
2{\rm Im}\Big[
\varphi_k(\tau)
\varphi^\star_k(\tau_j)
\Big]
\\&
\hspace{0.5in}
\int\frac{d^3p}{(2\pi)^3}
2{\rm Im}\Big[
\varphi_k(\tau)
\varphi^\star_k(\tau_{i-})
\varphi_{|{\bm k}-{\bm p}|}(\tau_j)
\varphi^\star_{|{\bm k}-{\bm p}|}(\tau_{i-})
\varphi_p(\tau_j)
\varphi^\star_p(\tau_{i-})
\Big],
\numberthis
\end{align*}
where $\tau_{i-}=\tau_i(1+i\omega)$ captures the effect of the $i\omega$ prescription. The time integrals in the gravitational wave graphs involving self-interactions can be written using the analytical model in a similar manner.

Let us examine the convergence of the momentum integrals in these two diagrams. In the IR limit $p\rightarrow 0$, the modes $\varphi_p$ are frozen on superhorizon scales and are therefore time-independent, so we have, for the first integral,
\begin{equation}
r(k)\equiv
\int\frac{d^3p}{(2\pi)^3}
2{\rm Im}\Big[
\varphi_{|{\bm k}-{\bm p}|}(\tau)
\varphi^\star_{|{\bm k}-{\bm p}|}(\tau_i)
\varphi_k(\tau)
\varphi^\star_k(\tau_i)
\varphi_p(\tau)
\varphi^\star_p(\tau_i)
\Big]
\propto
\int\frac{d^3p}{(2\pi)^3}
|\varphi_p(\tau)|^2.
\end{equation}
In the IR limit the spectrum is flat, so that $|\varphi_p|^2\propto 1/p^3$ and the integrand scales as $dp/p$. Using an IR cutoff $\Lambda_{\rm IR}$, we find
\begin{equation}
r(k)=f(k)+g(k)\log\bigg(\frac{k}{\Lambda_{\rm IR}}\bigg),
\end{equation}
for two functions $f(k)$ and $g(k)$. Following \cite{Ballesteros:2024zdp}, we assume that IR divergences are unphysical and drop out of observable quantities, so we do not consider them here. Since we do not know the exact form of $g(k)$, the finite piece $f(k)$ can be extracted numerically by performing the integral twice with two different cutoffs $\Lambda_{\rm IR}$ and $\Lambda_{\rm IR}^\star$, chosen sufficiently far so that the integrand has reached its asymptotic $dp/p$ form, and using
\begin{equation}
f(k)
=
\bigg[
\log\bigg(
\frac{k}{\Lambda_{\rm IR}^\star}
\bigg)r(k)+
\log\bigg(\frac{\Lambda_{\rm IR}}{k}\bigg)r_\star(k)
\bigg]
\log\bigg(
\frac{\Lambda_{\rm IR}}{\Lambda_{\rm IR}^\star}\bigg)^{-1}.
\end{equation}
The integral in the second diagram can be dealt with in exactly the same way.

In the UV limit $p\rightarrow \infty$, convergence is ensured by the $i\omega$ prescription. To see this, note that in this limit modes are oscillating in the Bunch-Davies vacuum,
\begin{equation}
\varphi_p=-\frac{e^{-ip\tau}}{2M_p a \sqrt{k\epsilon}}.
\label{eq:i_epsilon}
\end{equation}
We therefore have, for the second integral \cite{Ballesteros:2024zdp}
\begin{align*}
\ell_{ij}(k)
&=
\int\frac{d^3p}{(2\pi)^3}
\varphi_{|{\bm k}-{\bm p}|}(\tau_j)
\varphi^\star_{|{\bm k}-{\bm p}|}(\tau_{i-})
\varphi_p(\tau_j)
\varphi^\star_p(\tau_{i-})
\\&=
\frac{1}{4 a(\tau_j)^2\epsilon(\tau_j)}
\frac{1}{4 a(\tau_i)^2\epsilon(\tau_i)}
\int\frac{d^3p}{(2\pi)^3}
\frac{e^{-i(|{\bm k}-{\bm p}|+p)(\tau_j-\tau_{i-})}}{|{\bm k}-{\bm p}|p}
\\&=
\frac{1}{2\epsilon(\tau_i)a(\tau_i)^2}\frac{1}{2\epsilon(\tau_j)a(\tau_j)^2}
\frac{ie^{-ik(\tau_j-\tau_i)}}{16\pi^2(\tau_j-\tau_i)},\numberthis
\label{eq:iwpresc}
\end{align*}
where $\tau_-=\tau(1+i\omega)$ and we have used the $i\omega$ prescription to perform the integral in the last line, which is clearly oscillating otherwise. Numerically, the $i\omega$ prescription can be implemented by simply multiplying $\varphi_k\rightarrow \varphi_k e^{\omega k\tau}$, where $\varphi_k$ is the numerically-obtained solution to the Mukhanov-Sasaki equation, and $\omega$ is sufficiently small to not affect the solution on superhorizon scales, where $|k\tau|\ll 1$.

\begin{figure}
\centering
\includegraphics[width=0.49 \textwidth]{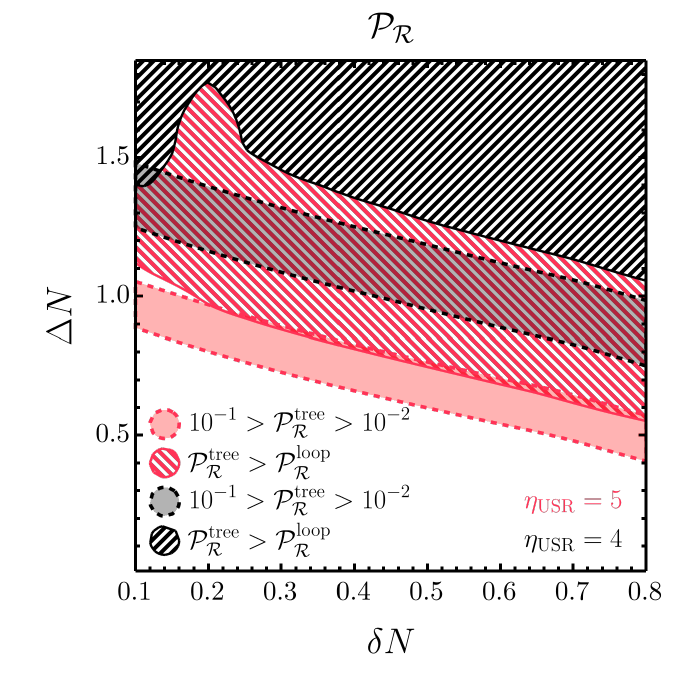} 
\includegraphics[width=0.49 \textwidth]{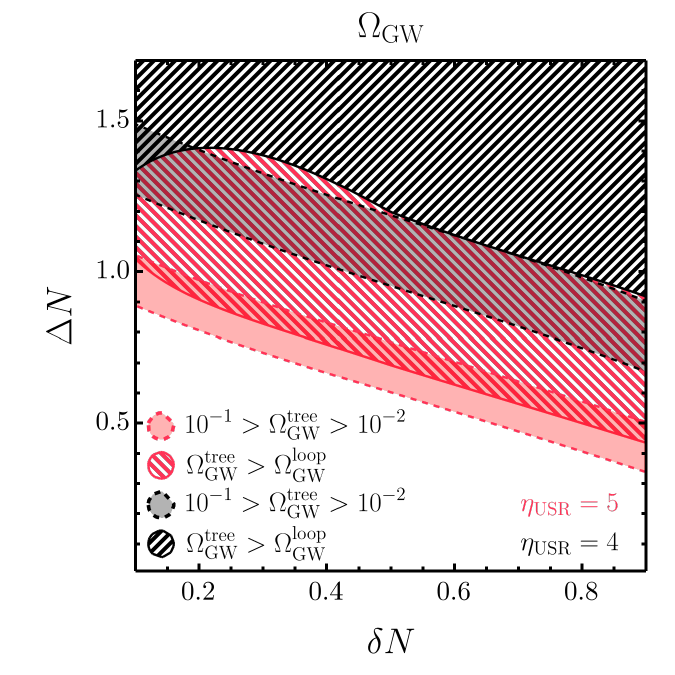} 
\caption{
\it Left panel: Scan over parameter space of the tree-level and one-loop contributions to the power spectrum for $\eta_{\rm CR}=-1$ and $\eta_{\rm USR}=4$ (black) and $\eta_{\rm CR}=-2$ and $\eta_{\rm USR}=5$ (red). The hatched region with solid boundary, obtained by comparing the tree-level spectrum at its peak to the one-loop correction, shows where both terms become comparable and perturbation theory breaks down. The shaded bands with dashed boundaries show the region in which the tree-level power spectrum is between $10^{-2}$ and $10^{-1}$. Right panel: Same as left panel, but comparing the leading (one-loop) contribution to $\Omega_{\rm GW}$ to the two-loop correction. The shaded bands denote the region in which the peak of the GW spectrum lies between $10^{-9}$ and $10^{-7}$.
}
\label{fig:pars}
\end{figure}

The situation for the gravitational wave integrals is much simpler. The diagrams in (\ref{eq:loop_prop}) converge once the divergences in the one-loop power spectrum are taken care of. On the other hand, similarly to the scalar power spectrum case we have just discussed, the integrals in eqs.\,(\ref{eq:c_diagram}) and (\ref{eq:z_diagram}) are divergent in both the IR and UV. We deal with these divergences by using cutoffs, effectively integrating only over the peak of the power spectrum, in the spirit of \cite{Li:2023xtl} (see also the discussion in footnote \ref{fn:divergences}). The convergence of the contributions in eqs.\,(\ref{eq:y_diagram}) and (\ref{eq:x_diagram}) in the UV is guaranteed by the $i\omega$ prescription. In the IR limit $p\rightarrow 0$, we have $I_k(p,|{\bm k}-{\bm p}|)\simeq {\rm const.}$ after averaging and $|\varphi_p|^2\propto 1/p^3$, so the entire integrand scales linearly with $p$ and therefore converges, and analogously for $q$. A similar analysis shows that the integral in eq.\,(\ref{eq:omega_ng}) for the gravitational wave anisotropies also converges.

\subsection{Numerical analysis}

The momentum integrals must be performed numerically. In the case of the one-loop power spectrum, once we fix the external momentum ${\bm k}$ to be aligned with the $\hat{\bm z}$ axis, none of the integrals depend on the azimuthal angle, so this angular integral can be immediately solved yielding a factor of $2\pi$. To perform the gravitational wave integrals, we use the following identities for the polarization tensors
\begin{align}
\Big[{\bm p}\cdot {\bm e}^s({\bm k})\cdot{\bm p}\Big]^2
&=
\frac{1}{2}p^4\sin^4\theta_p,\\
\Big[{\bm p}\cdot {\bm e}^s({\bm k})\cdot{\bm p}\Big]\Big[{\bm q}\cdot {\bm e}^s({\bm k})\cdot{\bm q}\Big]
&=
\frac{1}{2}p^2q^2\cos 2\phi_-\sin^2\theta_p\sin^2\theta_q,
\end{align}
where the external momentum ${\bm k}$ is once again aligned with the $\hat{\bm z}$ axis, $\theta_p$ and $\theta_q$ denote the polar angles of $p$ and $q$, respectively, and $\phi_-=\phi_p-\phi_q$ the difference in azimuthal angles. Note that the only other place in the integrals where the azimuthal angles show up is in the modulus
\begin{equation}
|{\bm p}+{\bm q}|^2
=
q^2+p^2+2pq(\cos\theta_p\cos\theta_q+\cos\phi_-\sin\theta_p\sin\theta_q),
\end{equation}
and this quantity once again depends only on the combination $\phi_p-\phi_q$. By changing variables in one of these angles to $\phi_-$ the other angular integral can then be performed immediately, yielding a factor of $2\pi$. The remaining five integrals must be done numerically.

\begin{figure}
\centering
\includegraphics[width=0.49 \textwidth]{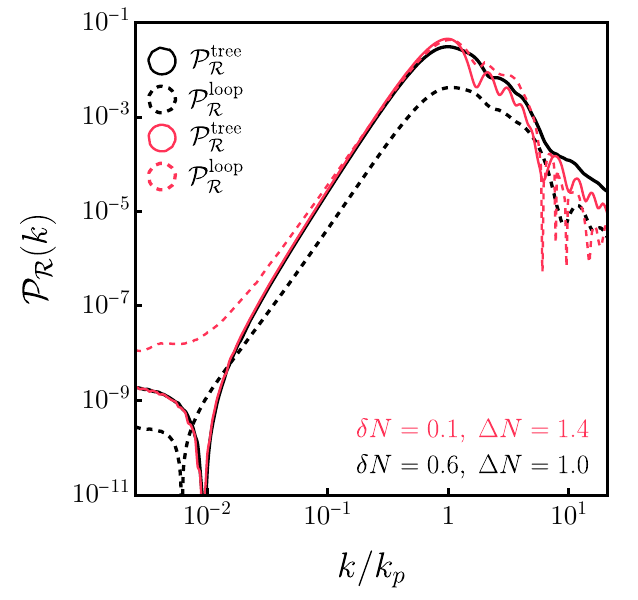} 
\includegraphics[width=0.49 \textwidth]{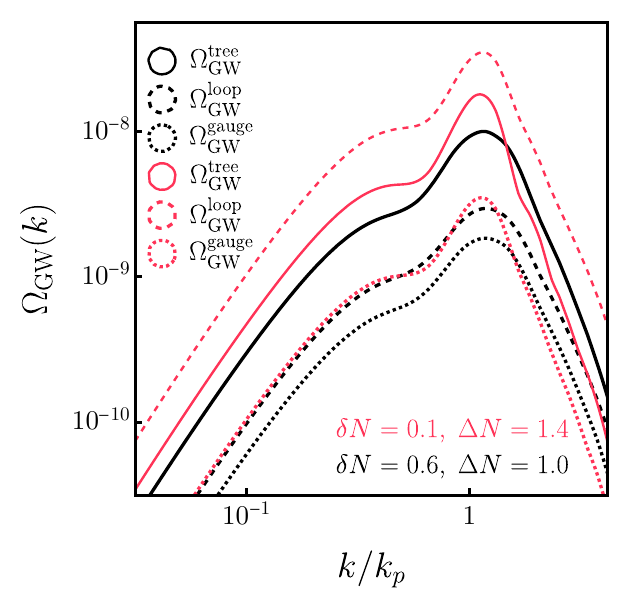} 
\caption{
\it Left panel: Tree-level scalar power spectra (solid) and loop corrections (dashed) for $\eta_{\rm USR}=4$ and $\eta_{\rm CR}=-1$. We set $\delta N=0.1$ and $\Delta N=1.4$ for the example in red, which breaks perturbation theory, and $\delta N=0.6$ and $\Delta N=1.0$ for the example in black, which does not. Right panel: Gravitational wave spectra corresponding to the examples in the left panel. The leading (one-loop) contribution corresponds to the solid lines, whereas the dashed lines denote the two-loop corrections. The dotted lines reflect the sum of the contributions of the diagrams  corresponding to replacing one of the propagator with the corrections in eqs.\,(\ref{eq:fg_loops_1}) and (\ref{eq:fg_loops_2}), together with (\ref{eq:c_diagram}) and (\ref{eq:z_diagram}), that is, this line corresponds to the effect of the gauge transformation only, which is subdominant with respect to the self-interaction diagrams for reasonable values of $\eta_{\rm CR}$.
}
\label{fig:one_loop}
\end{figure}

A scan of the entire parameter space is presented both for the one-loop power spectrum and the gravitational wave energy density in Fig.\,\ref{fig:pars} for realistic $\mathcal{O}(1)$ values of $\eta_{\rm CR}$, extending the results of \cite{Ballesteros:2024zdp}. We find that, for values of the power spectrum of order $10^{-1}$ and $\eta_{\rm CR}=-1$, perturbation theory breaks down only for nearly instantaneous transitions ($\delta N\lesssim 0.15$) in and out of ultra-slow-roll.\footnote{The hatched regions in the plot on the left panel of Fig.\,\ref{fig:pars} have a small notch toward the left. This bump has no particular physical meaning and is simply due to the fact that we have chosen the peak of the spectrum to compare the tree-level and one-loop contributions, and the loop correction oscillates in this zone (see e.g.\,the left panel of Fig.\,\ref{fig:one_loop}). The region looks slightly different depending on which value of $k$ is chosen for the comparison. In other words, for values near the notch in the left panel (e.g.\,for $\delta N = 0.2$ and $\Delta N = 1.6$), perturbation theory is in fact violated for values $k> k_{p}$. This artifact in the procedure exists only for a narrow range of values near the edge (for which perturbation theory is only borderline acceptable anyway) and thus its effect on the qualitative nature of the results is negligible.} Increasing $\eta_{\rm CR}$ has two effects. One is that the entire plot is shifted downwards, which is to be expected since now the spectrum grows and decays at a different rate.\footnote{Note that we choose different values of $\eta_{\rm USR}$ in such a way that the Wands duality holds \cite{Wands:1998yp}, see the caption of Fig.\,\ref{fig:pars}. It is in fact the increase in $\eta_{\rm USR}$ that reduces the duration of the USR phase, shifting the problematic region downwards.} The second effect is that the region where perturbation theory breaks is pushed much closer to the values of interest for the power spectrum. This occurs because increasing $\eta_{\rm CR}$ increases $f_{\rm NL}$ and $g_{\rm NL}$, so that diagrams involving these quantities have a larger effect and loop corrections become more relevant. We have chosen realistic values of $\eta_{\rm CR}$ (see e.g.\,the model in \cite{Ballesteros:2020qam}), and find that for larger values it might be impossible to generate a large power spectrum without breaking perturbation theory.\footnote{The fact that the nonlinearity corrections are subdominant if perturbation theory holds was also discussed in \cite{Atal:2019cdz}.} These conclusions are, unsurprisingly, essentially mirrored by the gravitational wave energy density.

Particular examples of the loop-corrected scalar spectrum and gravitational wave energy density are shown in Fig.\,\ref{fig:one_loop}. We have chosen two representative examples, one in which perturbation theory breaks down (in red) and one in which it does not (in black). We find that the main contribution to the gravitational wave energy density comes not from the $f_{\rm NL}$ and $g_{\rm NL}$ corrections (as long as reasonable values are chosen for $\eta_{\rm CR}$), but from the diagrams in (\ref{eq:loop_prop}) that involve replacing one scalar propagator by its loop-corrected value.

\begin{figure}
\centering
\includegraphics[width=0.49 \textwidth]{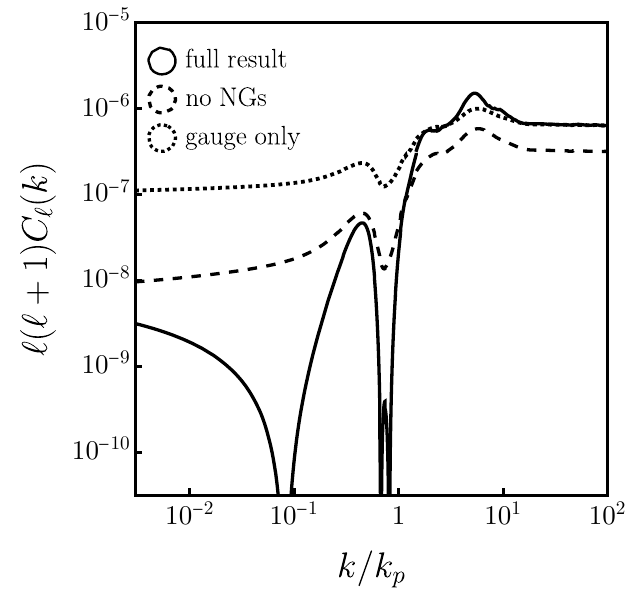} 
\caption{
\it Angular power spectrum $C_\ell$ from eq.\,(\ref{eq:anisotropies}) as a function of the wavenumber $k$ for the black example of Fig.\,\ref{fig:one_loop}. The dashed line corresponds to the angular coefficients obtained in the absence of non-Gaussianities, that is, the last term of eq.\,(\ref{eq:anisotropies}). The solid line corresponds to the full result. The dotted line was obtained by including only the $f_{\rm NL}$ term in eq.\,(\ref{eq:omega_ng}), i.e.\,neglecting self-interactions.
}
\label{fig:anis}
\end{figure}

Our results for the angular power spectrum, obtained via eq.\,(\ref{eq:omega_ng}), are shown in Fig.\,\ref{fig:anis}. As we have seen, loop corrections can only change $\Omega_{\rm GW}$ by an insignificant amount once we impose the constraint that perturbation theory should be obeyed. The relevance of the non-Gaussian corrections becomes apparent only when we consider anisotropies, since even a small amount of non-Gaussianity can significantly change the shape of the angular coefficients. For a stochastic background of primordial origin and in the absence of non-Gaussianities \cite{Dimastrogiovanni:2022eir}, anisotropies do not carry any more information than one would obtain by measuring the two-point function of gravitational waves,\footnote{A similar statement was made in \cite{Malhotra:2022ply} in a different context to that of non-Gaussianities. The authors showed that the effect on the anisotropies of an equation of state different to that of radiation in the early Universe is washed out, so that also in this setting all the physical information is already carried by the GW frequency spectrum.} since the last term of eq.\,(\ref{eq:anisotropies}) depends only on the tilt of $\Omega_{\rm GW}$. We therefore find that anisotropies are a smoking gun in assessing whether or not a primordial gravitational wave signal is induced by scalar modes.

\phantomsection
\section*{Conclusions}
\addcontentsline{toc}{section}{Conclusions}

By computing the inflationary correlators in the $\delta\phi$ gauge and transforming the result to the $\mathcal{R}$ gauge, we have determined the relevant interactions during an ultra-slow-roll phase of inflation (following \cite{Ballesteros:2024zdp}) and provided the Feynman rules to perform calculations diagrammatically. We have used this method to compute the non-Gaussian corrections to the energy density and anisotropies of the induced gravitational wave spectrum. We find that non-Gaussianities are, in general, not of the local form. If reasonable\,--\,that is, $\mathcal{O}(1)$\,--\,values are chosen for $\eta_{\rm CR}$, the diagrams involving $f_{\rm NL}$ and $g_{\rm NL}$ are essentially negligible, and self-interactions dominate, requiring use of the full machinery of the in-in formalism to perform the calculation. We have done this numerically by using the analytical model presented in \cite{Ballesteros:2024zdp}, which allows us to control the duration and smoothness of the transitions in and out of ultra-slow-roll.

We have calculated the one-loop scalar power spectrum and found that, in the presence of a constant-roll stage with a nonvanishing $\eta$ after the ultra-slow-roll phase, the region of parameter space in which perturbation theory breaks down expands significantly. In particular, for $\eta_{\rm CR}\lesssim -2$, we find that it becomes impossible to obtain a power spectrum of order $\mathcal{P}_\mathcal{R}\sim 10^{-1}$ and remain within the perturbative regime, even for smooth transitions. Unsurprisingly, we find that the induced gravitational wave spectrum inherits this property. These results are illustrated in Fig.\,\ref{fig:pars}.

We have determined the set of diagrams involved in the calculation of the induced gravitational wave spectrum at two loops, extending the results of \cite{Li:2023xtl}. We find fourteen independent diagrams, five of which vanish due to helicity conservation, and five of which consist on replacing one of the scalar propagators in the leading-order result by its loop-corrected value. The remaining diagrams involve computing five-dimensional integrals (after eliminating the time integrals using the aforementioned analytical model), which we have done numerically. Example spectra and their loop corrections are presented in Fig.\,\ref{fig:one_loop}, together with the result obtained by considering only some of the diagrams corresponding to the local piece of the non-Gaussianity for comparison. The Figure illustrates the fact that the latter is negligible and the self-interactions of the inflaton dominate.

Finally, we determined how non-Gaussianities arising from self-interactions affect the anisotropies of the induced gravitational wave spectrum. We have derived a simple expression (\ref{eq:omega_ng}) for the scalar-induced anisotropies in terms of the tilt of the scalar power spectrum on small scales, presented here for the first time. The angular coefficients are shown in Fig.\,\ref{fig:anis}. There is a clear difference between the full result and the one obtained by including only the non-Gaussianities of the local form arising from the nonlinear relation between $\delta\phi$ and $\mathcal{R}$, making anisotropies a key observable in determining whether a gravitational wave signal is induced by scalar modes or not. The examination of the prospects of measuring this anisotropy using space-based interferometers is left for future work.\footnote{The detection of anisotropies in the gravitational wave background is a delicate issue. Pulsar Timing Array searches, for instance, are subject to uncertainties due to cosmic variance, so that even isotropic backgrounds would yield anisotropic sky maps due to interference between different sources \cite{Konstandin:2024fyo}. Whether some of these issues can be ameliorated with space-based interferometers remains to be seen.} Let us remark, however, that in the case of both LISA \cite{Alonso:2020rar,LISACosmologyWorkingGroup:2022kbp} and PTAs \cite{NANOGrav:2023tcn} present estimates yield an observable value of $C_\ell\sim\mathcal{O}(1)$.\footnote{One must be careful to normalize the results in \cite{Alonso:2020rar,LISACosmologyWorkingGroup:2022kbp} by the monopole contribution $C_0\sim\Omega_{\rm GW}^2$} This is the order of magnitude of the results expected from astrophysical sources \cite{NANOGrav:2023tcn}, and detecting a possible inflationary signal therefore represents a significant challenge. Moreover, note that the approximation we have made in going from eq.\,(\ref{eq:spherical_bessel}) to eq.\,(\ref{eq:anisotropies}) is used precisely to keep only the leading contribution to the integral at low values of $\ell$, which are the ones we are most likely to observe in the near future \cite{Alonso:2020rar,LISACosmologyWorkingGroup:2022kbp,NANOGrav:2023tcn}, but keeping the full integral would yield a non-trivial dependence on $\ell$ for larger values. This dependence, in combination with the frequency dependence of the angular coefficients, could in principle be used to distinguish between different sources (see e.g.\,\cite{ValbusaDallArmi:2020ifo}).\footnote{We thank the anonymous referee for commenting on this point.} This is an intriguing possibility that would make for an interesting direction for future work.

Before closing, let us also comment briefly on the differences between the approach presented here and the calculation in the $\mathcal{R}$ gauge. Placing the scalar degree of freedom of the theory in the metric sector is convenient because it allows us to obtain the physically relevant variable $\mathcal{R}$ directly. However, this approach suffers from a number of disadvantages. The first is that it leads to a number of boundary terms in the action upon integration by parts which enter the loop corrections (since the in-in formalism is sensitive to the time boundary). These terms have introduced spurious results in the past and have been the subject of debate, see e.g.\,\cite{Franciolini:2023agm,Ballesteros:2024zdp,Braglia:2024zsl}. The second disadvantage is that the relevant interactions during a USR phase are more difficult to identify, since the terms arising from the inflaton potential and the gauge transformation mix with each other. The third is that the connection between the gauge transformation and the $\delta N$ formalism result in eq.\,(\ref{eq:delta_n_formalism}) is not as obvious, due to the aforementioned mixing. Finally, the fact that one must calculate the interaction Hamiltonian in the interaction picture has similarly led to spurious results when the prescription is not implemented correctly, see \cite{Franciolini:2023agm,Ballesteros:2024zdp}. In contrast, the calculation in the $\delta\phi$ gauge is completely transparent, and technically much simpler. A detailed comparison between the two calculations would nonetheless be an interesting direction for future work.

\section*{Acknowledgments}
We are grateful to Guillem Dom\`enech and Thomas Konstandin for comments on the draft, and to Guillermo Ballesteros and Jes\'us Gamb\'in Egea for multiple discussions about their work and the subtleties of the in-in formalism. This work is supported by the Deutsche Forschungsgemeinschaft under Germany’s Excellence Strategy – EXC 2121 Quantum Universe – 390833306.


\printbibliography

\end{document}